\documentclass{aa}
\usepackage[varg]{txfonts}
\usepackage{graphicx}
\usepackage{natbib}
\usepackage{color}
\usepackage[colorlinks=true,linkcolor=blue,citecolor=blue,urlcolor=blue]{hyperref}

\bibpunct{(}{)}{;}{a}{}{,} 

\begin{document}
\title{X-ray radiative transfer in protoplanetary disks}
\subtitle{The role of dust and X-ray background fields}
\author{Ch. Rab\inst{1,3}
  \and M. Güdel\inst{1}
  \and P. Woitke\inst{2}
  \and I. Kamp\inst{3}
  \and W.-F. Thi\inst{4}
  \and M. Min\inst{5,6}
  \and G. Aresu\inst{7}
  \and R. Meijerink\inst{8}}
\institute{University of Vienna, Dept. of Astrophysics, T\"urkenschanzstr. 17, 1180 Wien, Austria 
         \email{christian.rab@univie.ac.at}  
\and SUPA, School of Physics \& Astronomy, University of St. Andrews, North Haugh, St. Andrews KY16 9SS, UK
\and Kapteyn Astronomical Institute, University of Groningen, P.O. Box 800, 9700 AV Groningen, The Netherlands
\and Max-Planck-Institut für extraterrestrische Physik, Giessenbachstrasse 1, 85748 Garching, Germany
\and SRON Netherlands Institute for Space Research, Sorbonnelaan 2, 3584 CA Utrecht, The Netherlands
\and Astronomical institute Anton Pannekoek, University of Amsterdam, Science Park 904, 1098 XH, Amsterdam, The Netherlands
\and INAF, Osservatorio Astronomico di Cagliari, via della Scienza 5, 09047 Selargius, Italy 
\and Leiden Observatory, Leiden University, PO Box 9513, 2300 RA Leiden, The Netherlands}
\date{Received 26 June 2017 / Accepted 16 November 2017}
\abstract
{The X-ray luminosities of T Tauri stars are about two to four orders of magnitude higher than the luminosity of the contemporary Sun. As these stars are born in clusters, their disks are not only irradiated by their parent star but also by an X-ray background field produced by the cluster members.}
{We aim to quantify the impact of X-ray background fields produced by young embedded clusters on the chemical structure of disks. Further we want to investigate the importance of the dust for X-ray radiative transfer in disks.}
{We present a new X-ray radiative transfer module for the radiation thermo-chemical disk code P{\tiny RO}D{\tiny I}M{\tiny O} (PROtoplanetary DIsk MOdel), which includes X-ray scattering and absorption by both the gas and dust component. The X-ray dust opacities can be calculated for various dust compositions and dust size distributions. For the X-ray radiative transfer we consider irradiation by the star and by X-ray background fields. To study the impact of X-rays on the chemical structure of disks we use the well established disk ionization tracers N$_2$H$^+$ and HCO$^+$.}
{For evolved dust populations (e.g. grain growth), X-ray opacities are mostly dominated by the gas; only for photon energies $E\gtrsim5-10\,\mathrm{keV}$, dust opacities become relevant. Consequently the local disk X-ray radiation field is only affected in dense regions close to the disk midplane. X-ray background fields can dominate the local X-ray disk ionization rate for disk radii $r\gtrsim20\,\mathrm{au}$. However, the N$_2$H$^+$ and HCO$^+$ column densities are only significantly affected in case of low cosmic-ray ionization rates ($\lesssim10^{-19}\,\mathrm{s^{-1}}$), or if the background flux is at least a factor of ten higher than the flux level of $\approx10^{-5}\,\mathrm{erg\,cm^{-2}s^{-1}}$ expected for clusters typical for the solar vicinity.}
{Observable signatures of X-ray background fields in low-mass star-formation regions, like Taurus, are only expected for cluster members  experiencing a strong X-ray background field (e.g. due to their location within the cluster). For the majority of the cluster members, the X-ray background field has only little impact on the disk chemical structure.}

\makeatletter
\renewcommand*\aa@pageof{, page \thepage{} of \pageref*{LastPage}}
\makeatother
\maketitle
\section{Introduction}
Strong X-ray emission is a common property of pre-main sequence stars. \mbox{T Tauri} stars, often considered as young solar analogs, show strong X-ray emission with luminosities in the range of approximately $10^{29}-10^{31}\, \mathrm{~erg\,s^{-1}}$ \citep[e.g.][]{Preibisch2005,Gudel2007d}, which is about $10^2-10^4$ times higher than the X-ray luminosity of the contemporary Sun \citep{Feigelson2002a}. The origin of such high X-ray luminosities is likely the enhanced stellar and magnetic activity of the young stars \citep[e.g][]{Feigelson2002a}, but also jets close to the star and the protoplanetary disk \citep{Gudel2007e} caused by interaction of the stellar and disk magnetic fields \citep[e.g. X-wind][]{Shu1997q} might contribute. Accretion shocks probably do not contribute significantly to the X-ray emission of T~Tauri stars \citep{Gudel2007f}, but accreting material absorbs soft X-rays and might cool the hot coronal gas \citep{Gudel2007g}.

X-ray irradiation plays an important role for the thermal and chemical structure of protoplanetary disks. Soft X-rays heat the upper disk layers to temperatures larger than $5000\mathrm{~K}$ \citep{Glassgold2004,Nomura2007ah,Aresu2011} and possibly drive, together with far and extreme ultraviolet radiation, disk photo-evaporation \citep[e.g.][]{Ercolano2008c,Gorti2009}. Diagnostics of the interaction of X-rays with the disk atmosphere are mainly atomic lines \citep[e.g.][]{Gorti2004,Meijerink2008,Ercolano2008c,Adamkovics2011,Aresu2012a}. These lines trace the hot upper layers (vertical column densities of $10^{19}-10^{20}\,\mathrm{cm^{-2}}$) in the inner $\approx 10-50\,\mathrm{au}$ of protoplanetary disks \citep{Glassgold2007,Aresu2012a}. X-rays can influence atomic line emission via heating (e.g. [OI], \citealt{Aresu2014}) and/or direct ionization (e.g. the neon ion fine-structure lines \citealt{Glassgold2007}). \citet{Gudel2010a} found a correlation between the [NeII] $12.81\,\mathrm{\mu m}$ line, observed with the \emph{Spitzer} Space telescope, and stellar X-ray luminosity in a sample of 92 pre-main sequence stars. Such a correlation is consistent with predictions of several thermo-chemical disk models \citep{Meijerink2008,Gorti2008a,Schisano2010,Aresu2012a}.

Hard X-ray emission with energies larger than $1\,\mathrm{keV}$ can also penetrate deeper disk layers where they become an important ionization source of molecular hydrogen \citep[e.g.][]{Igea1999,Ercolano2013b} and therefore drive molecular-ion chemistry. However, in those deep layers X-rays compete with other high energy ionization sources like cosmic rays, decay of short-lived radionuclides \citep[e.g.][]{Umebayashi2009,Cleeves2013c} and stellar energetic particles \mbox{\citep{Rab2017}}. Observationally those ionization processes can be traced by molecular ions, where HCO$^+$ and N$_2$H$^+$ are the most frequently observed ones \citep[e.g.][]{Thi2004a,Dutrey2007a,Dutrey2014,Oberg2011f,Cleeves2015a,Guilloteau2016}. 
In contrast to the atomic lines, X-ray heating does not play a prominent role for molecular ion line emission. Consequently molecular ions are good tracers of chemical processes such as ionization. Nevertheless, there is no clear picture yet, both observationally and theoretically, about the main ionization process determining the abundances of those molecules. For example, \citet{Salter2011c} found no correlation of HCO$^+$ millimetre line fluxes with stellar properties like mass, bolometric luminosity or X-ray luminosity. 

Predictions from models concerning the impact of X-ray emission on HCO$^+$ and N$_2$H$^+$ are quite different. The models of \citet{Teague2015b} indicate a strong sensitivity of the HCO$^+$ column density to the X-ray luminosity at all disk radii assuming an ISM like cosmic-ray ionization rate. However, in the models of \citet{Cleeves2014a} HCO$^+$ and N$_2$H$^+$ column densities become sensitive to stellar X-rays only if the cosmic-ray ionization rate is as low as $\zeta_\mathrm{CR}\approx10^{-19}\,\mathrm{s^{-1}}$. \citet{Walsh2012} concluded that far-UV photochemistry plays a more dominant role for molecular ions than X-rays (using $\zeta_\mathrm{CR}\approx10^{-17}\,\mathrm{s^{-1}}$). \citet{Rab2017} included energetic stellar particles as additional high-energy ionization source. In their models N$_2$H$^+$ is sensitive to X-rays but only for low cosmic-ray ionization rates, where HCO$^+$ might be dominated by stellar particle ionization, assuming that the paths of the particles are not strongly affected by magnetic fields that may guide them away from the disk.

An aspect not yet considered in radiation thermo-chemical disk models are X-ray background fields of embedded clusters. \citet{Adams2012} estimated the X-ray background flux distribution for typical clusters in the solar vicinity. They  find that the background flux impinging on the disk surface can be higher than the stellar X-ray flux in the outer disk regions ($r\gtrsim10\,\mathrm{au}$).  

In this work we introduce a new X-ray radiative transfer module for the radiation thermo-chemical disk code P{\tiny RO}D{\tiny I}M{\tiny O}. This module includes X-ray scattering and a detailed treatment of X-ray dust opacities, considering different dust compositions and grain size distributions. In addition, we also include an X-ray background field, as proposed by \citet{Adams2012}, as additional disk irradiation source. We investigate the impact of X-ray background fields on the disk chemistry in particular on the common disk ionization tracers HCO$^+$ and N$_2$H$^+$.

In Sect.~\ref{sec:methods}, we describe the X-ray radiative transfer module and our disk model used to investigate the impact of stellar and interstellar X-ray radiation. Our results are presented in Sect.~\ref{sec:results}. At first we show the resulting X-ray disk ionization rates for models including scattering, X-ray dust opacities and X-ray background fields. The impact on the disk ion chemistry is studied via comparison of HCO$^+$ and N$_2$H$^+$ column densities. In Sect.~\ref{sec:discussion}, we discuss observational implications of X-ray background fields also in context of enhanced UV background fields. A summary and our main conclusions are presented in Sect.~\ref{sec:conclusions}.
\section{Methods}
\label{sec:methods}
We use the radiation thermo-chemical disk code P{\tiny RO}D{\tiny I}M{\tiny O} (PROtoplanetary DIsk MOdel) to model the thermal and chemical structure of a passive disk irradiated by the stellar and interstellar radiation fields. P{\tiny RO}D{\tiny I}M{\tiny O} solves consistently for the dust temperature, gas temperature and chemical abundances in the disk \citep{Woitke2009a} and includes modules producing observables like spectral energy distributions \citep{Thi2011} and line emission \citep{Kamp2010,Woitke2011}. 

The disk model we use here is based on the so-called reference model developed for the DIANA\footnote{DIANA website: \url{http://diana-project.com/}} (DiscAnalysis) project. This model is consistent with typical dust and gas observational properties of \mbox{T Tauri} disks, and is described in detail in \citet{Woitke2016} and \citet{Kamp2017}. We therefore provide here only a brief overview of this reference model (Sect.~\ref{sec:diskrefmodel}). For this work we also use different dust size distributions to study the impact of dust on the X-ray RT; those models are described in Sect.~\ref{sec:modelgroups}. The new X-ray radiative transfer module of P{\tiny RO}D{\tiny I}M{\tiny O} is described in Sect.~\ref{sec:xrayrt}. 
\subsection{Reference model}
\label{sec:diskrefmodel}
In the following we describe the gas and dust disk structure of the reference model and the chemical network we used. In Table~\ref{table:discmodel} we provide an overview of all model parameters including the properties of the central star. The stellar and interstellar X-ray properties are described in Sect.~\ref{sec:xrayrt}.
\begin{table}
\caption{Main parameters for the reference disk model.}
\label{table:discmodel}
\centering
\begin{tabular}{l|c|c}
\hline\hline
Quantity & Symbol & Value  \\
\hline
stellar mass                          & $M_\mathrm{*}$                    & $0.7~M_\sun$\\
stellar effective temp.               & $T_{\mathrm{*}}$                  & 4000~K\\
stellar luminosity                    & $L_{\mathrm{*}}$                  & $1.0~L_\sun$\\
FUV excess                            & $L_{\mathrm{FUV}}/L_{\mathrm{*}}$ & 0.01\\
FUV power law index                   & $p_{\mathrm{UV}}$                 & 1.3\\
X-ray luminosity                      & $L_\mathrm{X}$                    & $10^{30}~\mathrm{erg\,s^{-1}}$\\
X-ray emission temp.                  & $T_{\mathrm{X}}$                  & $2\times10^7$ K\\
\hline
strength of interst. FUV              & $\chi^\mathrm{ISM}$               & 1\tablefootmark{a}\\
\hline
disk gas mass                         & $M_{\mathrm{disk}}$               & $0.01~M_\sun$\\
dust/gas mass ratio                   & $d/g$                             & 0.01\\
inner disk radius                     & $R_{\mathrm{in}}$                 & 0.07~au\\
tapering-off radius                   & $R_{\mathrm{tap}}$                & 100~au\\
column density power ind.             & $\epsilon$                        & 1.0\\

reference scale height                & $H(100\;\mathrm{au})$             & 10~au\\
flaring power index                   & $\beta$                           & 1.15\\
\hline
min. dust particle radius             & $a_\mathrm{min}$                  & $\mathrm{0.05~\mu m}$\\
max. dust particle radius             & $a_\mathrm{max}$                  & 3 mm\\
dust size dist. power index           & $a_\mathrm{pow}$                  & 3.5\\
turbulent mixing param.               & $\alpha_{\mathrm{settle}}$        & $10^{-2}$\\
max. hollow volume ratio\tablefootmark{b}              & $V_{\mathrm{hollow,max}}$         & 0.8\\
dust composition                      & {\small Mg$_{0.7}$Fe$_{0.3}$SiO$_3$}  & 60\%\\
(volume fractions)                    & {\small amorph. carbon}                    & 15\%\\
& {\small porosity}                          & 25\%\\
\hline
PAH abun. rel. to ISM                 & $f_\mathrm{PAH}$                  & 0.01\\
chem. heating efficiency              & $\gamma^\mathrm{chem}$            & 0.2\\
\hline
\end{tabular}
\tablefoot{
If not noted otherwise, these parameters are kept fixed for all our models presented in this work. For more details on the parameter definitions see \citet{Woitke2009a,Woitke2011,Woitke2016}.
\tablefoottext{a}{$\chi^\mathrm{ISM}$ is given in units of the Draine field \citep{Draine1996b,Woitke2009a}.}
\tablefoottext{b}{We use distributed hollow spheres for the dust opacity calculations \citep{Min2005,Min2016}.}
}
\end{table}
\subsubsection{Gas disk structure}
\label{sec:gasdisk}
\begin{figure}
\centering
\includegraphics{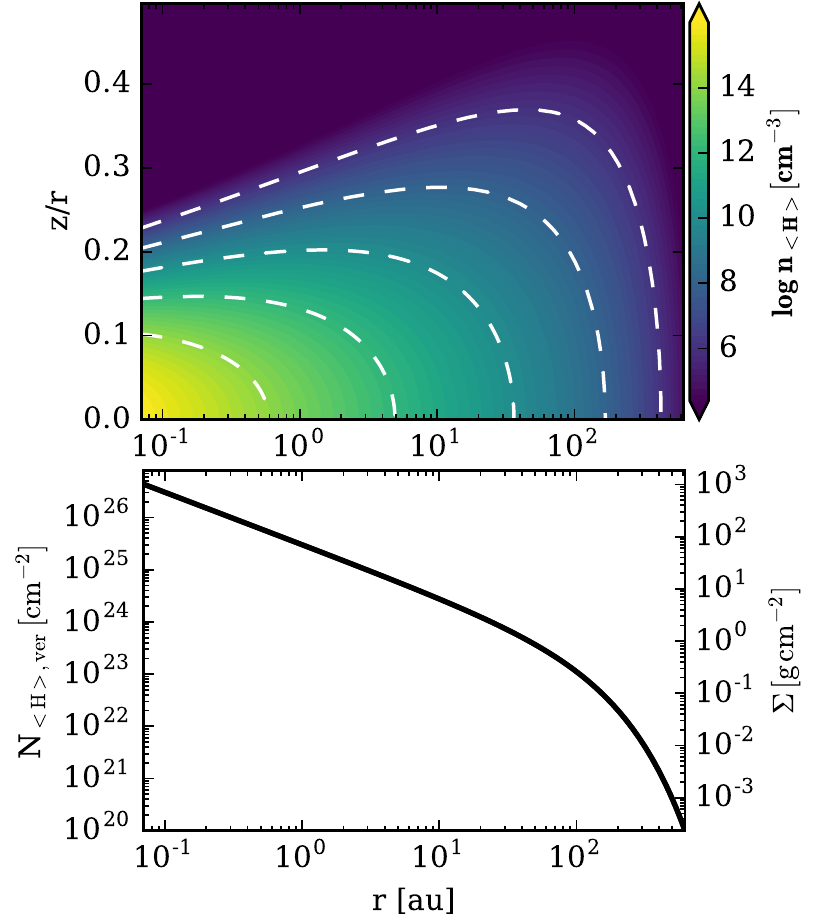}
\caption{Gas disk structure. The top panel shows the total hydrogen number density $n_\mathrm{\langle H\rangle}$. The height of the disk $z$ is scaled by the radius $r$. The white dashed contours correspond to the density levels shown in the colourbar. The bottom panel shows the total vertical hydrogen column number density $N_\mathrm{\langle H\rangle,ver}$ as a function of radius where on the right hand side also the scale for the surface density $\Sigma$ in $\mathrm{g\,cm^{-2}}$ is given.}
\label{fig:diskstruc}
\end{figure}

We use a fixed parameterized density structure for the disk. 
The axissymmetric flared 2D gas density structure as a function of the cylindrical coordinates $r$ and $z$ (height of the disk) is given by \citep[e.g.][]{Lynden-Bell1974b,Andrews2009a,Woitke2016} 
\begin{equation}
  \label{eqn:density}
    \rho(r,z)=\frac{\Sigma(r)}{\sqrt{2\pi}\cdot h(r)}\exp\left(-\frac{z^2}{2h(r)^2}\right)\;\;\mathrm{[g\,cm^{-3}]}\;, 
\end{equation} 
For the vertical disk scale height $h(r)$ we use a radial power-law  
\begin{equation}
h(r)=H\mathrm{(100\;au)}\left(\frac{r}{100\;\mathrm{au}}\right)^{\beta}
\end{equation} 
where $H\mathrm{(100\;au)}$ is the disk scale height at $r=100\,\mathrm{au}$ (here $H\mathrm{(100\;au)}=10\,\mathrm{au}$) and $\beta=1.15$ is the flaring power index. For the radial surface density we use again a power-law with a tapered outer edge  
\begin{equation}
\Sigma(r)=\Sigma_0 \left(\frac{r}{R_\mathrm{in}}\right)^{-\epsilon}\exp\left(-\left(\frac{r}{R_{\mathrm{tap}}}\right)^{2-\epsilon}\right)\;\;\mathrm{[g\,cm^{-2}]}\;. 
\label{eqn:surfdens}
\end{equation}
The inner disk radius is $R_\mathrm{in}=0.07\,\mathrm{au}$ (the dust condensation radius), the characteristic radius is $R_\mathrm{tap}=100\,\mathrm{au}$ and the outer radius is $R_\mathrm{out}=620\,\mathrm{au}$ where the total vertical hydrogen column density is as low as $N_\mathrm{\langle H \rangle,ver} \approx 10^{20}\,\mathrm{cm^{-2}}$. The constant $\Sigma_0$ is given by the disk mass $M_\mathrm{disk}=0.01\,M_\sun$ and  determined via the relation $M_\mathrm{disk}=2\pi\int\Sigma(r)r\mathrm{dr}$ to be $1011\,\mathrm{[g\,cm^{-2}]}$. The 2D gas density structure and the radial column density profile of the disk are shown in Fig.~\ref{fig:diskstruc}. The gas density structure is the same for all models presented in this paper.
\subsubsection{Dust disk structure}
\label{sec:dustdisk}
We assume a dust to gas mass ratio of $\delta=0.01$, which also determines the total dust mass. However, due to dust evolution processes like dust growth, dust settling and drift \citep[see e.g.][]{Birnstiel2012}, the dust density structure in a protoplanetary disk does not necessarily follow the gas density structure. 

We account for this by including a dust size distribution with a minimum dust grain size of $a_\mathrm{min}=0.05\,\mathrm{\mu m}$ and a maximum grain size of $a_\mathrm{max}=3000\,\mathrm{\mu m}$. The size distribution itself is given by a simple power-law $f(a)\propto a^{-a_\mathrm{pow}}$ with $a_\mathrm{pow}=3.5$ \citep{Mathis1977b}. Dust settling is incorporated by applying the method of \citet{Dubrulle1995}, using a turbulent mixing parameter of $\alpha_\mathrm{settle}=10^{-2}$. This results in a grain size and gas density dependent dust scale height and the dust to gas mass ratio varies within the disk. For example at $r=100\,\mathrm{au}$ the local dust to gas mass ratio varies from $\delta_\mathrm{local}\approx0.1$ close to the midplane to values $\delta_\mathrm{local}\lesssim0.001$ in the upper layers of the disk. We note that the total dust to gas mass ratio $\delta$ stays the same (for more details see \citealt{Woitke2016}). We use porous grains composed of a mixture of amorphous carbon and silicate (see Table~\ref{table:discmodel}). The resulting dust grain density is $\rho_\mathrm{gr}=2.1\,\mathrm{g\,cm^{-3}}$. Our model does not account for possible radial drift of large dust particles (see Sect.~\ref{sec:dustevol} for a discussion).

In P{\tiny RO}D{\tiny I}M{\tiny O} the same dust model is consistently used for the radiative transfer, including X-rays and the chemistry. For more details on the dust model and opacity calculations see \citet{Woitke2016,Min2016}.
\subsubsection{Chemical network}
\label{sec:chemistry}
Our chemical network is based on the gas-phase chemical database UMIST~2012 \citep{McElroy2013b} where we only use a subset of reactions according to our selection of species. Additionally to the gas phase reactions from the UMIST database we include detailed X-ray chemistry \citep{Meijerink2012a}, charge exchange chemistry of PAHs (Polycyclic aromatic hydrocarbons, \citealt{Thi2014}, \citealt{Thi2017}), excited H$_2$ chemistry, $\mathrm{H}_2$ formation using the analytical function of \citet{Cazaux2002,Cazaux2004} and adsorption and thermal, photo and cosmic-ray desorption of ices (including PAHs). Dust surface chemistry is not included in our model.

The chemical network used here is described in detail in \citet{Kamp2017}. We used their so called large chemical network which consists of $235$ chemical species ($64 $ of them are ices) and $3143$ chemical reactions. The element abundances are listed in \citet{Kamp2017}. The element abundances correspond to the group of low metal abundances \citep[e.g][]{Graedel1982c,Lee1998}. Further details concerning the chemistry of $\mathrm{HCO^+}$ and $\mathrm{N_2H^+}$ and the used binding energies are given in \citet{Rab2017}.

To solve for the chemical abundances we used the steady-state approach. In \citet{Woitke2016} and \citet{Rab2017} comparisons of time-dependent and steady-state chemistry models are presented, which show that in our models steady-state is reached within typical lifetimes of disks in most regions of the disk. In \citet{Rab2017} it is shown that the assumption of steady-state is well justified for $\mathrm{HCO^+}$ and $\mathrm{N_2H^+}$ (see also \citealt{Aikawa2015}). The resulting differences between the steady-state and time-dependent models in the radial column density profiles of the molecular ions are not significant for our study.
\subsection{Model groups}
\label{sec:modelgroups}
\begin{table}
\caption{Dust models.}
\label{table:models}
\centering
\begin{tabular}{llcc}
\hline\hline
Name & Parameters & $A_\mathrm{V}=1$\tablefootmark{a} & Surface \tablefootmark{b} \\
     &            & $(\mathrm{cm^{-2}})$& $(\mathrm{cm^{2})}$ \\ 
\hline
small grains  & single size &  & \\ 
SG & $a=0.1\,\mathrm{\mu m}$ &$1.4(21)$\tablefootmark{c}& $2.8(-21)$\\  
& no settling & & \\
\hline
medium grains  & $a_\mathrm{min}=0.005\,\mathrm{\mu m}$ & &  \\
MG &  $a_\mathrm{max}=1000\,\mathrm{\mu m}$ &$1.9(22)$& $7.6(-22)$ \\
&  $a_\mathrm{pow}=3.7$ & & \\
\hline
large grains &$a_\mathrm{min}=0.05\,\mathrm{\mu m}$& & \\
LG, reference &$a_\mathrm{max}=3000\,\mathrm{\mu m}$&$1.2(23)$&$2.7(-23)$\\ 
 & $a_\mathrm{pow}=3.5 $ & & \\
\hline
\end{tabular}
\tablefoot{
\tablefoottext{a}{hydrogen column density $N_\mathrm{\langle H\rangle}$ where the visual extinction $A_\mathrm{V}$ is unity.}
\tablefoottext{b}{total dust surface per hydrogen nucleus (unsettled value)}
\tablefoottext{c}{$a(b)$ means $a\times10^b$.}
}
\end{table} 
\begin{figure}
\resizebox{\hsize}{!}{\includegraphics{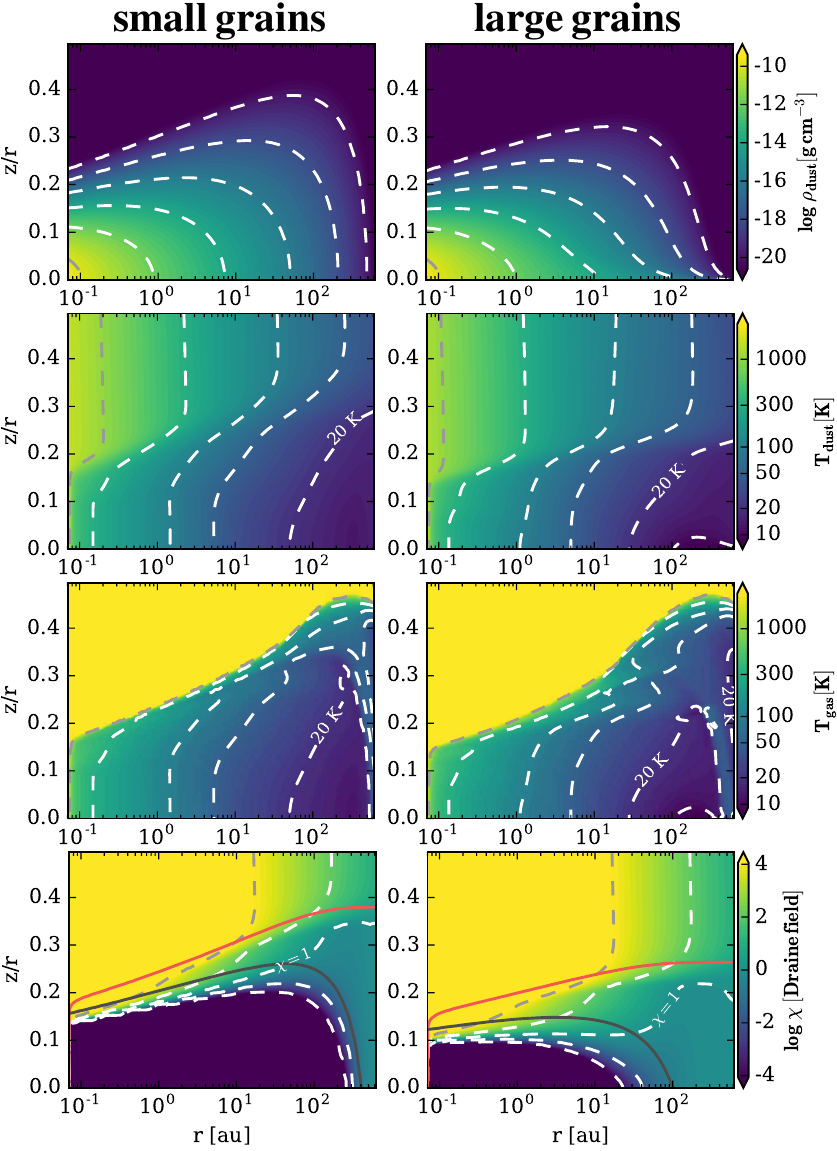}}
\caption{Large grains (left column) and small grains (right column) disk model. From top to bottom the dust density, dust temperature, gas temperature and the disk UV radiation field $\chi$ in units of the Draine field is shown. The dashed contours in each plot correspond to the levels shown in the respective colourbar. The additional contours in the UV plots (bottom row) indicate where the radial (red solid line) and vertical (black solid line) visual extinction are equal to unity. For both models the same gas density structure as shown in Fig.~\ref{fig:diskstruc} is used.}
\label{fig:diskstruc_dust}
\end{figure}
\subsubsection{Dust models}
\label{sec:dustmodels}
For our investigations of the impact of the dust on the X-ray radiative transfer and on the molecular column densities we use three different dust size distributions. All three distributions have the same dust composition as listed in Table~\ref{table:discmodel}. The parameters varied are $a_\mathrm{min}$, $a_\mathrm{max}$ and $a_\mathrm{pow}$. In all three dust models the gas density structure is the same as described in Sect.~\ref{sec:gasdisk} and the dust to gas mass ratio is $\delta=0.01$. The details of the dust models are provided in Table~\ref{table:models}.

The small grains model (SG) includes only a single dust size with  $a=0.1\,\mathrm{\mu m}$ and no dust settling. Although such a dust model is likely not a good representation for the conditions in protoplanetary disks, is is useful as a reference and to show the impact of the dust on the X-ray radiative transfer and chemical disk structure.

The medium grain and large grain models are more appropriate in the context of grain growth and dust settling. (see Sect.~\ref{sec:dustdisk}). Although, such dust models are a simplified representation of dust evolution in disks \citep[][]{Birnstiel2012,Vasyunin2011,Akimkin2013,Facchini2017}, they still provide insight on the role of dust for the chemistry and X-ray radiative transfer. The main difference of the medium and large grain model is the amount of small particles. In the medium grain model about $10\%$ of the total dust mass is in grains with $a\leq 1\,\mathrm{\mu m}$ whereas in the large grain models it is only about $1.5\%$. Both models include dust settling as described in Sect.~\ref{sec:dustdisk}.

Most relevant for the chemistry is the total dust surface area per hydrogen nucleus (i.e. for the freeze-out of molecules). The dust surface area varies by about two orders of magnitude in our models (Table~\ref{table:models}, see also \citealt{Woitke2016,Rab2017a} for details).

In Fig.~\ref{fig:diskstruc_dust} we show the dust density structure, the dust and gas temperature and the local UV radiation field in the disk, for the small grains and large grains model. The main difference in the dust models is the visual extinction $A_\mathrm{V}$, which significantly affects the dust temperature structure and the local disk radiation field. 
\subsubsection{Cosmic rays}
\label{sec:crmodels}
There is some uncertainty about how many of the cosmic rays actually reach the disk of T Tauri stars. Similar to the Sun, the stellar wind of young stars, which might be significantly stronger compared to the Sun, can power a heliosphere-like analog which is called a ``T~Tauriosphere''. The existence of such a T~Tauriosphere might reduce the cosmic-ray ionization rate in the disk by several orders of magnitude \citep{Cleeves2013}.  

To account for this, we use two different cosmic-ray input spectra, one is the canonical local ISM cosmic-ray spectrum \citep{Webber1998c} and the second is a modulated spectrum which accounts for the suppression of cosmic-rays by a heliosphere. For the latter we use the ``Solar Max'' spectrum of \citealt{Cleeves2013}. To calculate the cosmic-ray ionization rate we use the fitting formula of \citet{Padovani2013c} and \citet{Cleeves2013}. The ISM cosmic-ray spectrum gives a cosmic-ray ionization rate per hydrogen nucleus of $\zeta_\mathrm{CR}\approx10^{-17}$ and the Solar Max spectrum gives $\zeta_\mathrm{CR}\approx10^{-19}$, which is consistent with the upper limit of the total $\mathrm{H}_2$ ionization rate in \object{TW Hya} derived by \citet{Cleeves2015a}. We call these two model groups ``ISM cosmic rays'' and ``low cosmic rays'', respectively.

\subsection{X-ray radiative transfer}
\label{sec:xrayrt}
For the X-ray radiative transfer (RT) we extended the already available radiative transfer module of P{\tiny RO}D{\tiny I}M{\tiny O} to the X-ray wavelength regime. The 2D radiative transfer problem including scattering is solved with a ray-based method and a simple iterative scheme ($\Lambda$-iteration). The radiative transfer equation is solved for a coarse grid of wavelengths bands. For each band the relevant quantities (e.g. incident intensities, opacities) are averaged over the wavelength range covered by each band. The details of this method are described in \citet{Woitke2009a}. For the X-ray regime we find that about 20 wavelength bands are sufficient to represent the energy range of $0.1-20\,\mathrm{keV}$ used in our models. Besides the stellar radiation also interstellar radiation fields (UV and X-rays) are considered. For the interstellar radiation fields we assume that the disk is irradiated isotropically.

The X-ray RT module provides the X-ray radiation field for each point in the disk as a function of wavelength. Those values are used in the already available X-ray chemistry module of P{\tiny RO}D{\tiny I}M{\tiny O} to calculate the X-ray ionization rate for the various chemical species. The X-ray chemistry used in this paper is the same as presented in \citet{Aresu2011} and \citet{Meijerink2012a}. As the chemistry influences the gas composition which in turn determines the gas opacities, we iterate between the X-ray RT and the chemistry. For each X-ray radiative transfer step the chemical abundances are kept fixed where for each chemistry step the X-ray ionization rates are fixed. We find that typically three to five iterations are required until convergence is reached. 
\subsubsection{X-ray scattering}
\label{sec:xscattering}
Compton scattering, which reduces to Thomson scattering at low energies, is the dominant scattering process in the X-ray regime. The anisotropic behaviour of Compton scattering is treated via an approximation by reducing the isotropic scattering cross-section by a factor $(1-g)$, where $g=\langle\cos\theta\rangle$ is the asymmetry parameter and $\theta$ is the scattering angle \citep[see also][]{Laor1993}. We use this approach for both the gas and the dust scattering cross-sections. We call this reduced cross-section the pseudo anisotropic (pa) scattering cross-section. $g$ is zero for isotropic scattering and approaches unity in case of strong forward scattering. We apply this approach because the treatment of anisotropic scattering in a ray-based radiative transfer code is expensive, in contrast to Monte Carlo radiative transfer codes. Although this method is a simple approximation it is very efficient and we find that our results concerning X-ray gas radiative transfer and the X-ray ionization rate are in reasonably good  agreement with the Monte Carlo X-ray radiative transfer code \texttt{MOCASSIN} (see Appendix.~\ref{sec:comperco}).
\subsubsection{X-ray gas opacities}
\label{sec:gasopac}
For the X-ray gas opacities we used the open source library \texttt{xraylib}\footnote{xraylib source code \url{https://github.com/tschoonj/xraylib}} \citep{Brunetti2004g,Schoonjans2011a}. This library provides the X-ray absorption and scattering cross-sections (Rayleigh and Compton scattering) for atomic and molecular species. Concerning the X-ray gas absorption cross-section we find that the cross-section provided by \texttt{xraylib} are similar to the commonly used X-ray absorption cross-section of \citet{Verner1995d,Verner1996}. 

For the calculation of the X-ray gas opacities we used the same low-metal element abundances as are used for the chemistry (see Sect.~\ref{sec:chemistry}). As a consequence of the strong depletion of heavy metals (\element{Na}, \element{Mg}, \element{Si}, \element{S}, \element{Fe}) by factors of 100 to 1000 compared to solar abundances, most of the absorption edges at higher energies ($E_\mathrm{X}>1\,\mathrm{keV}$) disappear and the gas phase opacities in this regime are mostly dominated by hydrogen and helium (see also \citealt{Bethell2011a}).

We note that we do not treat the depletion of gas-phase element abundances and the dust composition, which determines the element abundances in solids, consistently. Our approach here is to use the same dust properties (i.e. grain sizes, opacities) for the whole wavelength regime (X-ray to mm) considered in the radiative transfer modelling (see also Sect.~\ref{sec:dustopac} and Appendix~\ref{sec:xcscomp}). The chosen dust opacities are well suited for modelling of protoplanetary disks as discussed
in \citet{Woitke2016}. However, for the dust compositions and the dust to gas mass ratio used in this work the sum of the gas and dust X-ray opacities are consistent within a factor of two with pure \mbox{gas-phase} 
\mbox{X-ray} opacities using the solar element abundances of \mbox{\citet{Lodders2003}}.
\begin{figure}
\resizebox{0.95\hsize}{!}{\includegraphics{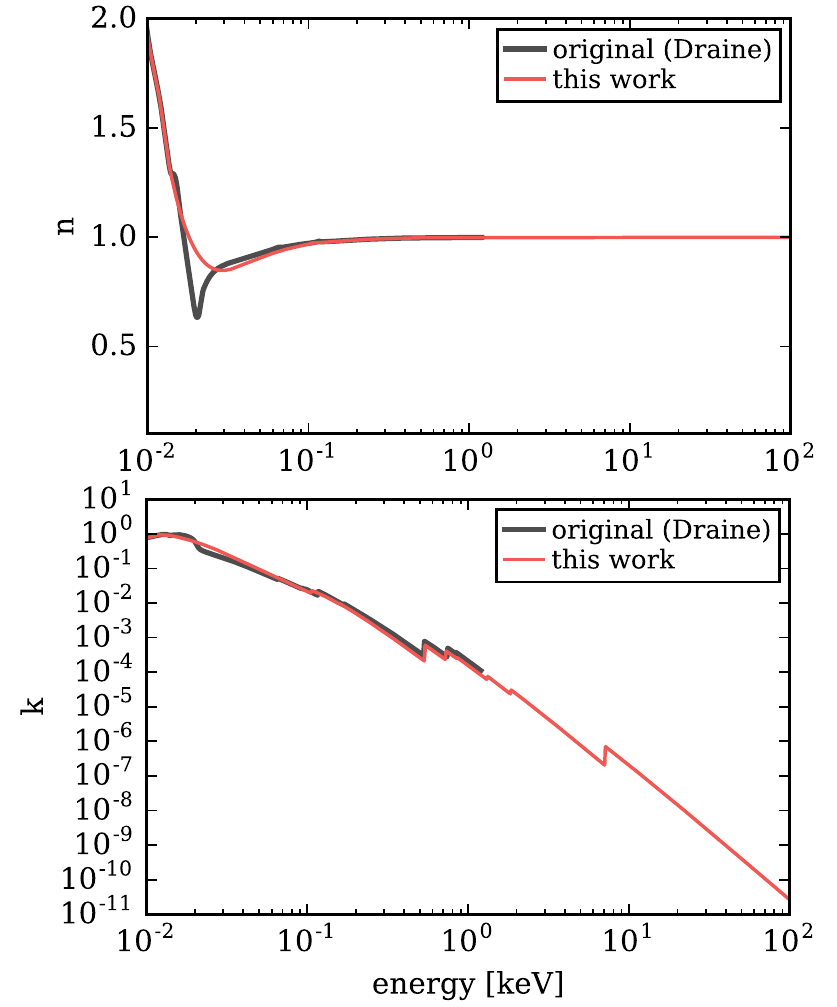}}
\caption{Optical constants $(n,k)$ for Astronomical Silicates in the X-ray regime. $n$ (top panel) is the real and $k$ (bottom panel) the imaginary part of the complex refractive index. The red solid lines show the results from this work, the black solid lines show the results from \citet{Draine2003b}.}
\label{fig:xdustconst}
\end{figure}
\begin{figure*}
\resizebox{\hsize}{!}{\includegraphics{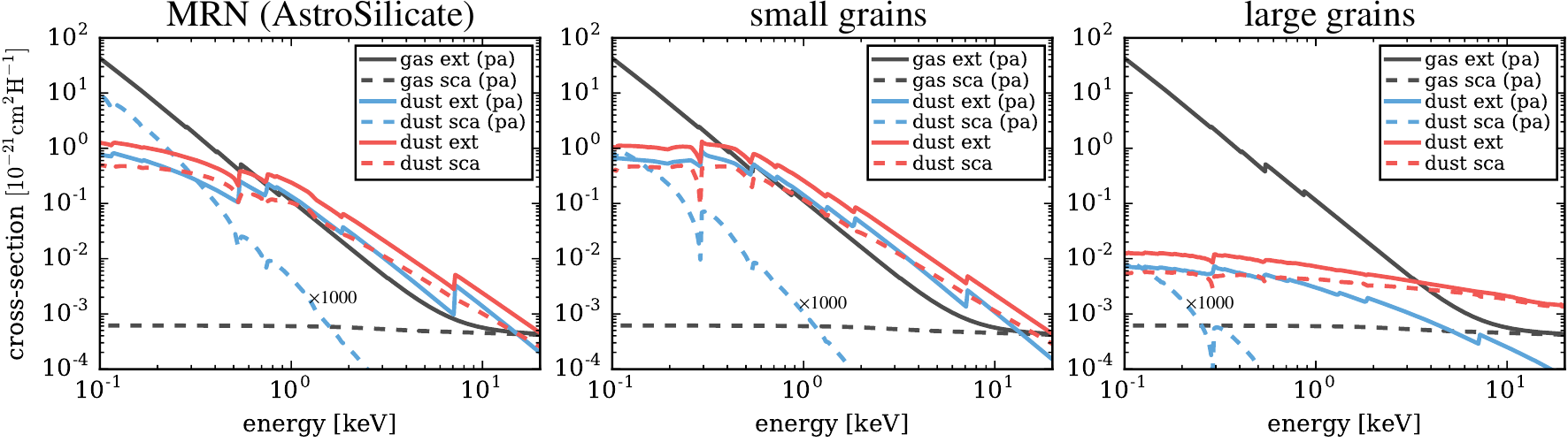}}
\caption{X-ray cross-sections for the gas (black lines) and dust (blue and red lines) per hydrogen nucleus for three different dust size distributions. The solid lines show the extinction (absorption+scattering) cross section, the dashed lines are  the scattering cross sections. The left panel is for the MRN size distribution \citep{Mathis1977b} using pure Astronomical Silicates \citep{Draine2003b}. The two other panels are for the small and large grains dust models (see Sect.~\ref{sec:dustmodels}). For the dust, the red lines are for isotropic and the blue lines for pseudo anisotropic (pa) scattering (see Sect.~\ref{sec:gasopac}). The pseudo anisotropic scattering values (blue dashed lines) are multiplied by a factor of 1000. The shown gas cross-sections are the same in all three panels.}
\label{fig:xdustopac}
\end{figure*}
\subsubsection{X-ray dust opacities}
\label{sec:dustopac}
In the literature, only X-ray optical constants for Astronomical Silicates and carbon are available \citep{Draine2003b}. However, for protoplanetary disks often a mixture of different dust species is used.

We implemented the method of \citet{Draine2003b} to calculate \mbox{X-ray} optical constants for various dust compositions. This method uses the available optical constants, or more precisely the dielectric function, for the ultraviolet to the millimetre wavelength range and additionally the atomic gas phase photo-electric cross-sections for the X-ray regime. With this, the imaginary part of the complex dielectric function can be constructed from the X-ray to the millimetre regime. Via the Kramers-Kronig relation the real part of the dielectric function can be calculated knowing the imaginary part. 

To be consistent with \citet{Draine2003b} we use the photo-electric cross-sections from \citet{Verner1996} to construct the imaginary part of the dielectric function. However, in contrast to \citet{Draine2003b} we do not include any additional measured data for the K edge absorption profiles (e.g. for graphite). The details of the absorption edges are less important here, as we are mainly interested in the resulting X-ray ionization rate which is an energy-integrated quantity.

In Fig.~\ref{fig:xdustconst} we compare the optical constants for Astronomical Silicates (MgFeSiO$_4$ composition) calculated via the above described method to the original optical constants provided by \citet{Draine2003b}. The deviations of our results from the \citet{Draine2003b} data are likely a consequence of our simplified treatment of absorption edges. However, our approach is sufficient for deriving X-ray ionization rates.

By using these newly available optical constants we calculated the X-ray dust opacities with a combination of the Mie-theory, the Rayleigh-Gans approximation \citep{Kruegel2002} and geometrical optics \citep{Zhou2003nn} similar to \citet{Draine2003b} and \citet{Ercolano2008c}. Also here we use the pseudo anisotropic scattering cross-sections in our calculations (see Sect.~\ref{sec:xscattering}).

In Fig.~\ref{fig:xdustopac} we show the dust extinction (sum of absorption and scattering cross-sections) and the scattering cross-sections per hydrogen nucleus for three examples of dust size distributions and compositions, in a similar way as presented in \citet{Draine2003b}. For comparison we also show the gas cross-section for the initial elemental abundances used in our models (low metal abundances, see \citealt{Kamp2017}), assuming that all hydrogen is molecular and all other elements are present as neutral atoms. For all cases shown in Fig.~\ref{fig:xdustopac} we assumed a gas to dust mass ratio of $0.01$.

The first panel in Fig.~\ref{fig:xdustopac} shows the results for an MRN (Mathis, Rumpl, Nordsieck; \citealt{Mathis1977b}) size distribution ($a_\mathrm{min}=0.005\,\mathrm{\mu m}$, $a_\mathrm{max}=0.25\,\mathrm{\mu m}$ and $a_\mathrm{pow}=3.5$) for pure Astronomical Silicates \citep{Draine2003b}. Although we do not use such a dust size distribution in our disk model the results are shown for reference. A comparison with Fig.~6 of \citet{Draine2003b} shows that our results are in good qualitative agreement with their results. However, \citet{Draine2003b} considered two individual dust populations with different dust compositions and size distribution: Astronomical Silicates and very small carbonaceous grains, the latter are not considered here. As a consequence the carbon absorption edge at around $0.3\,\mathrm{eV}$ is missing in our MRN dust model.

Fig.~\ref{fig:xdustopac} clearly shows that for energies below $1\,\mathrm{keV}$ the gas is the main opacity source even for the case of small grains. For higher energies dust becomes the dominant opacity source in the X-ray regime. However, considering a dust size distribution more typical for protoplanetary disks ($a_\mathrm{min}=0.05\,\mathrm{\mu m}$, $a_\mathrm{max}=3000\,\mathrm{\mu m}$ and $a_\mathrm{pow}=3.5$)‚ the dust extinction becomes only important for energies $\gtrsim 4\mathrm{keV}$. However, Fig.~\ref{fig:xdustopac} also shows that the dust scattering cross-section is significantly reduced if the g factor approximation is applied. The scattering phase function for dust is strongly forward peaked resulting in a g factor very close to unity (i.e. small scattering angle \citealt{Draine2003b}). The main consequence is that dust scattering is insignificant for disks as it does not produce a diffuse radiation field as most photons are simply forward scattered \citep{Bethell2011a}. 

The gas scattering cross-section is nearly independent of energy and becomes the dominant opacity source for energies $E_\mathrm{X}\gtrsim5\,\mathrm{keV}$. The scattering phase function of the gas is not strongly forward peaked as X-ray photons are mainly scattered by the electrons bound to hydrogen or helium. However, there is a slight decrease in the scattering cross-section with energy as Compton scattering becomes more anisotropic for higher energies, and in our simplified model this results in a reduced scattering cross-section.

We note that in our model the dust is simply an additional opacity source and we neglect actually any real interaction of X-rays with the dust, such as ionization or heating. The interaction of X-rays with solids is a very complex process \citep[see e.g.][]{Dwek1996d}. Besides heating \citep[e.g.][]{Laor1993} and ionization \citep{Weingartner2001e}, recent experiments indicate also possible dust amorphization by soft and hard X-rays \citep{Ciaravella2016,Gavilan2016}. However, we focus here on the X-ray ionization of the gas component and a detailed investigation of the impact of X-rays on the dust component is out of the scope of this paper.

Our chemical model also includes PAHs (see Sect.~\ref{sec:chemistry}). In Appendix~\ref{sec:xpahcs} we show that the expected X-ray absorption cross-sections for PAHs are too low to play a significant role in the attenuation of X-ray radiation. As our focus here is on the X-ray ionization rates a detailed modelling of the interaction of X-rays with PAHs is not considered.
\subsubsection{Stellar X-rays}
  To model the X-ray emission of \mbox{T Tauri} stars we assume that the origin of the emission is close to the stellar surface \citep[see e.g.][]{Ercolano2009b} and place the source of the emission on the star. The spectral shape of the emission is modeled with an isothermal bremsstrahlung spectrum \citep{Glassgold2009,Aresu2011} of the form
\begin{equation}
  \label{eqn:xspec}
  I(E)\propto\frac{1}{E}\cdot \exp\left(-\frac{E}{kT_\mathrm{X}}\right).      
\end{equation}
Where $E$ is the energy in keV, $I$ is the intensity, $k$ is the Boltzman constant and $T_\mathrm{X}=2\times10^7\,\mathrm{K}$ is the plasma temperature. We considered a X-ray energy range of $0.1-20\,\mathrm{keV}$ for the stellar spectrum. The spectrum is normalized to a given total X-ray luminosity $L_\mathrm{X}$ in the range of $0.3-10\,\mathrm{keV}$ as such an energy range is typical for reported observed X-ray luminosities \citep[e.g.][]{Gudel2007d}.

We note that it is also possible to use a more realistic thermal line plus continuum X-ray spectrum as input in P{\tiny RO}D{\tiny I}M{\tiny O} \citep[see][]{Woitke2016}. However, as we do not model here a particular source we use Eq.~\ref{eqn:xspec}, which is a reasonable approximation for the general shape of observed X-ray spectra \citep{Glassgold2009,Woitke2016}.
\subsubsection{X-ray background field}
\label{sec:XBGF}
A star embedded in a young cluster likely receives X-ray radiation from the other cluster members. In \citet{Adams2012} typical flux values for such a cluster X-ray background field (XBGF) are derived, where the values depend on the cluster properties (e.g. number of cluster members) and the position of the considered target within the cluster. Such a cluster background field is probably one to two orders of magnitude stronger than the diffuse extragalactic background field \citep{Adams2012}.

In \citet{Adams2012} only the total flux or energy averaged values of the XBGF are considered and now detailed X-ray radiative transfer method is applied. Here we included the XBGF in the energy dependent X-ray radiative transfer module. To do this we modelled the spectrum of the background field in the same way as the stellar X-ray spectrum. We used an isothermal bremsstrahlung spectrum (Eq.~(\ref{eqn:xspec})) with $T_\mathrm{X}=2\times10^7\,\mathrm{K}$ ($kT_\mathrm{X}\approx1.7\,\mathrm{keV}$) and normalized the spectrum to the given total flux in the energy range of $0.1-20\,\mathrm{keV}$. We assumed that the disk is irradiated isotropically by the XBGF. 

\citet{Adams2012} estimated a characteristic flux level for a cluster X-ray background field (XBGF) of $F_\mathrm{XBGF}=1-6\times10^{-5}\mathrm{\,erg\,cm^{-2}s^{-1}}$. As discussed by \citet{Adams2012} variations from cluster to cluster and also between single cluster members (i.e. location within the cluster) can be significant. Therefore we consider here flux levels for the XBGF in the range of $F_\mathrm{X}=2\times10^{-6}-2\times10^{-4}\mathrm{\,erg\,cm^{-2}s^{-1}}$, including the benchmark value of $2\times10^{-5}$ from \citet{Adams2012}. These values roughly cover the width of the X-ray background flux distributions derived by \citet{Adams2012}. We note that we considered here a slightly wider X-ray energy range as \citet{Adams2012}, who used $0.2-15\,\mathrm{keV}$. Therefore the given total flux levels differ slightly, for example $F_\mathrm{X}(0.1-20\,\mathrm{keV})=2\times10^{-5}\mathrm{\,erg\,cm^{-2}s^{-1}}$ corresponds to $F_\mathrm{X}(0.2-15\,\mathrm{keV})\approx1.4\times10^{-5}\mathrm{\,erg\,cm^{-2}s^{-1}}$. 

\citet{Adams2012} assumes that there is no absorption of X-rays within the cluster. However, absorption of  soft X-rays ($E\lesssim1\,\mathrm{keV}$) is possible, either by material between the star and the disk (see \citealt{Ercolano2009b}) or the interstellar medium itself. To account for such a scenario in a simple way we also used input spectra with low-energy cut-offs of $0.3$ and $1\,\mathrm{keV}$, respectively (see Sect.~\ref{sec:XBGFresults}).  

\begin{figure}
\centering
\resizebox{0.9\hsize}{!}{\includegraphics{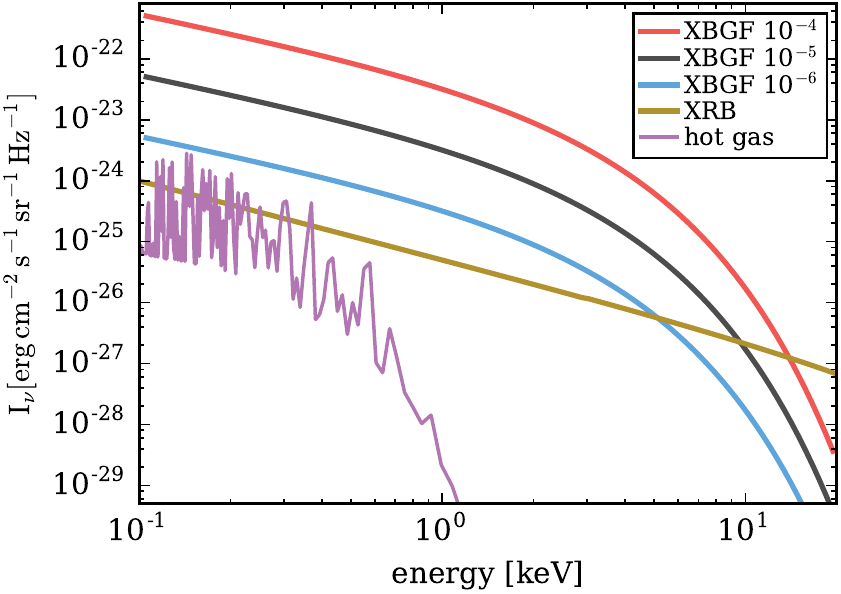}}
\caption{X-ray background field spectra. The black solid line shows a cluster X-ray background field as proposed by \citet{Adams2012} with a flux of $2\times10^{-5}~\mathrm{erg\,cm^{-2}\,s^{-1}}$ modelled with a bremsstrahlungs spectrum with $T_\mathrm{X}=2\times 10^7~$K. The red and blue solid lines are for fluxes ten times higher and lower, respectively. For comparison we also show the diffuse extragalactic X-ray background field (XRB, brown solid line) using the fits described in \citet{Fabian1992i} and a hot gas spectrum (magenta line) from e.g.  Supernova remnants (see \citealt{Tielens2005} chap. 1, \citealt{Slavin2008}).}
  \label{fig:BgSpectrum}
\end{figure}
In Fig.~\ref{fig:BgSpectrum} we show our XBGF spectra and additionally the spectrum measured for the diffuse extragalactic background field \citep{Fabian1992i} and a ``hot gas'' spectrum (e.g. produced by supernova remnants \citealt{Tielens2005}). This figure shows that typically the cluster XBGF will dominate the X-ray background flux impinging on the disk.
%
%
\section{Results}
\label{sec:results}
Our results are presented in the following way. In Sect.~\ref{sec:xionizationrates} we show the impact of scattering, dust opacities and X-ray background fields on the \mbox{X-ray} disk ionization rate $\zeta_\mathrm{X}$. In Sect.~\ref{sec:molcd} we present the molecular column densities of $\mathrm{HCO^+}$ and $\mathrm{N_2H^+}$ for our three different dust models and for models with and without X-ray background fields.
\begin{figure}
\centering
\includegraphics{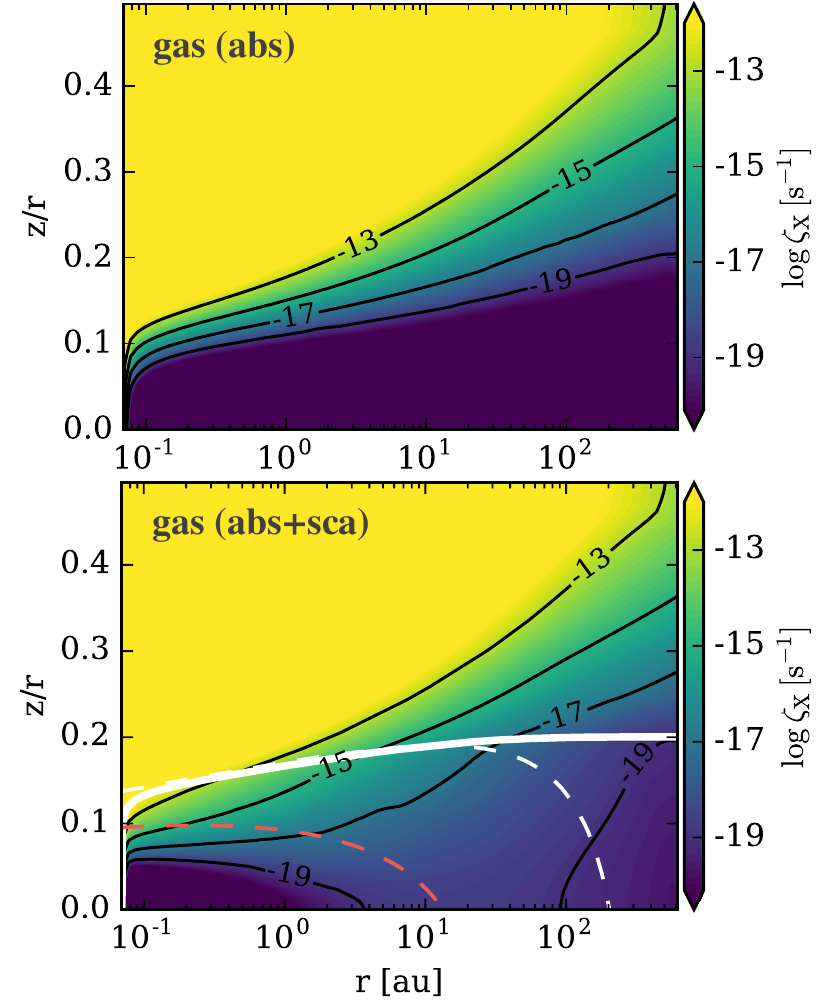}
\caption{X-ray ionization rate $\zeta_\mathrm{X}$ for the 2D disk structure. The top panel shows a model with only gas absorption the bottom panel a model with gas absorption and scattering. The white solid contour line shows $N_\mathrm{\langle H\rangle,rad}=2\times10^{24}\,\mathrm{cm^{-2}}$, which corresponds roughly to the X-ray scattering surface. Below this surface $\zeta_\mathrm{X}$ starts to be dominated by scattered X-ray photons. The white and red dashed lines show $N_\mathrm{\langle H\rangle,ver}=2\times10^{22}\,\mathrm{cm^{-2}}$ and $N_\mathrm{\langle H\rangle,ver}=2\times10^{24}\,\mathrm{cm^{-2}}$, respectively.} 
\label{fig:xionrateSca}
\end{figure}
\begin{figure*}
\centering
\includegraphics{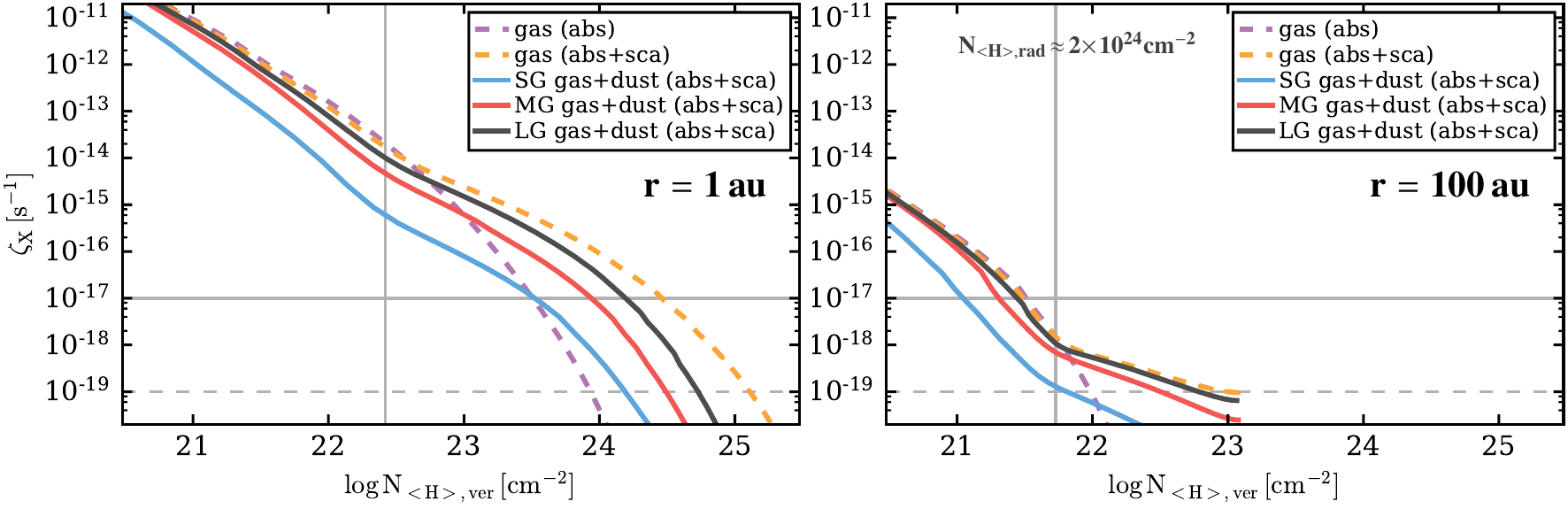}
\caption{X-ray ionization rate $\zeta_\mathrm{X}$ versus vertical column density $N_\mathrm{\langle H\rangle,ver}$ at disk radii of $1\,\mathrm{au}$ (left panel) and $100\,\mathrm{au}$ (right panel), respectively. The dashed lines show models with X-ray gas opacities only, where the purple line is for pure absorption and the orange line for absorption plus scattering. The solid lines are for models including X-ray dust opacities  (absorption+scattering), where results for the small grains (SG, blue), medium grains (MG, red) and large grains (LG, black) dust models are shown. The vertical grey line in both plots indicates the scattering surface at $N_\mathrm{\langle H\rangle,rad}=2\times10^{24}\,\mathrm{cm^{-2}}$ (see also Fig.~\ref{fig:xionrateSca}); at the right hand side of this line $\zeta_\mathrm{X}$ is dominated by scattered high energy photons. The horizontal lines mark the cosmic-ray ionization rates for the ISM ($\zeta_\mathrm{CR}\approx10^{-17}$), and low cosmic-ray case ($\zeta_\mathrm{CR}\approx10^{-19}\,\mathrm{s^{-1}}$).}
\label{fig:xionratesver}
\end{figure*}
\subsection{X-ray disk ionization rates}
\label{sec:xionizationrates}
In P{\tiny RO}D{\tiny I}M{\tiny O} the X-ray ionization rate is calculated individually for the single atoms and molecules (see \citealt{Meijerink2012a} for details). $\zeta_\mathrm{X}$ used in the following is simply the sum of the ionization rates of atomic and molecular hydrogen (we use a similar definition as \citealt{Adamkovics2011}). We note that we define $\zeta_\mathrm{X}$ per hydrogen nucleus, which is a factor of two lower compared to the ionization rate per molecular hydrogen. For all models discussed in this section $L_\mathrm{X}=10^{30}\,\mathrm{erg\,s^{-1}}$, if not noted otherwise.
\subsubsection{Impact of X-ray scattering}
X-ray scattering is already a quite common ingredient in protoplanetary disk modelling codes \citep[e.g.][]{Igea1999,Nomura2007ah,Ercolano2008c,Ercolano2013b,Cleeves2013}. In this section we want to compare our results concerning scattering to some of those models. We also briefly discuss the importance of X-ray scattering for $\zeta_\mathrm{X}$ considering both stellar X-rays and X-ray background fields.

\begin{figure*}
  \resizebox{\hsize}{!}{\includegraphics{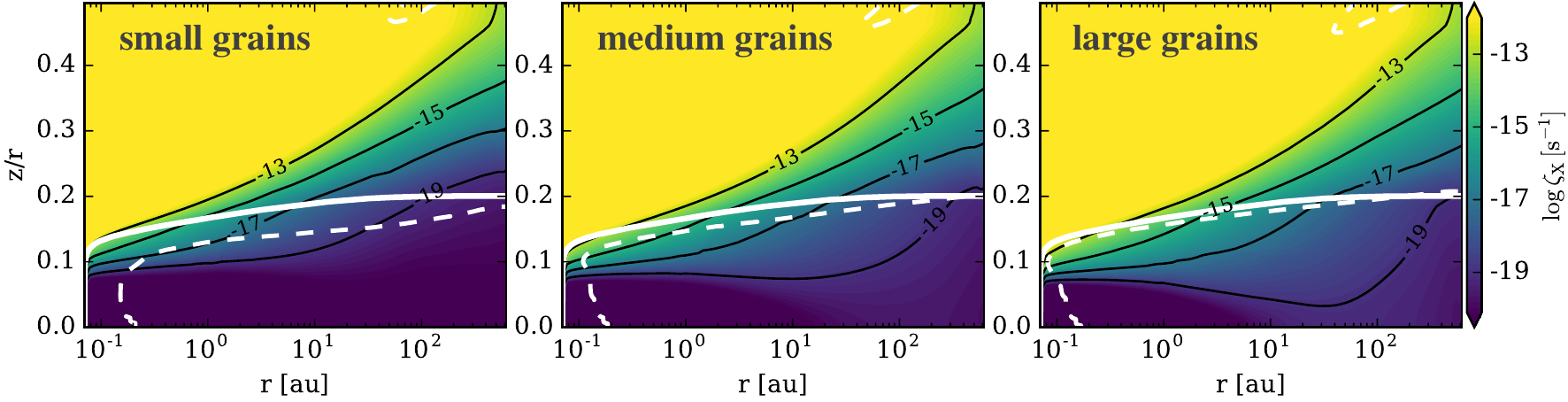}}
  \caption{X-ray ionization rate $\zeta_\mathrm{X}$ for the three different dust models including X-ray gas and dust opacities and scattering. The white solid contour shows $N_\mathrm{\langle H\rangle,rad}=2\times10^{24}\,\mathrm{cm^{-2}}$, which corresponds to the scattering surface. The dashed white contour shows where $\zeta_\mathrm{X}$ is equal to $\zeta_\mathrm{X,abs}$ of the gas absorption only model. Above this line $\zeta_\mathrm{X}\leq\zeta_\mathrm{X,abs}$ (additional absorption by the dust) below $\zeta_\mathrm{X}>\zeta_\mathrm{X,abs}$ (scattering).}
\label{fig:xionrates}
\end{figure*}
In Fig.~\ref{fig:xionrateSca} we show $\zeta_\mathrm{X}$ for the whole 2D disk structure for a model with X-ray gas absorption only and a model including scattering. In these models dust is not considered as an X-ray opacity source. The same models are also included in Fig.~\ref{fig:xionratesver} where $\zeta_\mathrm{X}$ is plotted as a function of vertical hydrogen column density $N_\mathrm{\langle H\rangle,ver}$ at radii of $1$ and $100\,\mathrm{au}$ distance from the central star.

As discussed already in \citet{Igea1999} and \citet{Ercolano2013b} it is possible to define a single scattering surface in the disk, because the scattering cross-section of the gas component is nearly constant over the whole X-ray energy range (see Fig.~\ref{fig:xdustopac}). A scattering optical depth of unity is reached at a hydrogen column density of $N_\mathrm{\langle H\rangle}\approx2\times10^{24}\,\mathrm{cm^{-2}}$ (see Fig.~\ref{fig:xdustopac} and \citealt{Igea1999}). For stellar X-rays the radial column density defines the location of the scattering surface in the disk (see Fig.~\ref{fig:xionrateSca}). Below the scattering surface the \mbox{X-ray} ionization rate is dominated by scattered X-rays. We want to note that the location of the scattering surface depends on the location of the \mbox{X-ray} source. For example in the models of \citet{Igea1999} and \citet{Ercolano2013b} the central \mbox{X-ray} source is located $12\,R_\sun$ above and below the star. Compared to our model where the \mbox{X-ray} source is the star itself, the scattering surface moves to deeper layers in the disk as the radial column density seen by stellar X-rays is reduced (see also Appendix~ \ref{sec:comperco}). This can cause differences in $\zeta_\mathrm{X}$ by about an order of magnitude in vertical layers close to the scattering surface \citep{Igea1999}. In terms of vertical column density the scattering surface is located at $N_\mathrm{\langle H\rangle,ver}\approx2\times10^{22}\,\mathrm{cm^{-2}}$ for $r\lesssim50\,\mathrm{au}$, but drops rapidly to lower vertical column densities due to the disk structure (see Fig.~\ref{fig:xionrateSca}).

An X-ray background field irradiates the disk isotropically. In Fig.~\ref{fig:xionrateSca} we also mark $N_\mathrm{\langle H\rangle,ver}\approx2\times10^{24}\,\mathrm{cm^{-2}}$ which roughly corresponds to the scattering surface for X-ray radiation entering the disk vertically (perpendicular to the midplane). This indicates that scattering is of less importance for X-ray background fields, as only in the region with $N_\mathrm{\langle H\rangle,ver}\gtrsim2\times10^{24}\,\mathrm{cm^{-2}}$ (i.e. r $\lesssim10\,\mathrm{au}$) scattering can become significant. However, in that region stellar X-rays will dominate the disk radiation field anyway (see Sect.~\ref{sec:xionbg}). Test models with and without X-ray scattering indeed showed that scattering is negligible for the case of an isotropic X-ray background radiation source. 

In Fig.~\ref{fig:xionratesver}, $\zeta_\mathrm{X}$ is also shown for models with and without scattering. This figure shows that at high column densities, $\zeta_\mathrm{X}$ is dominated by X-ray scattering, whereas at low column densities, $\zeta_\mathrm{X}$ is not significantly affected by scattering and is dominated by direct stellar X-rays. The reason is that the scattering cross-section becomes only comparable to the absorption cross-section for X-ray energies $E\gtrsim5\,\mathrm{keV}$ (see Fig.\ref{fig:xdustopac}). This means that mainly the energetic X-ray photons are scattered towards the midplane of the disk, where above the scattering surface $\zeta_\mathrm{X}$ is dominated by the softer X-rays which are not efficiently scattered. This is consistent with other X-ray radiative transfer models \citep[e.g.][]{Igea1999,Ercolano2013b,Cleeves2013c}.
\subsubsection{Impact of dust opacities}
In Fig.~\ref{fig:xionrates} we show the X-ray ionization rate for models with three different dust size distributions: small grains, medium grains and large grains (see Sect.~\ref{sec:dustmodels}). In these models both the gas and dust opacities are considered in the X-ray RT, but the X-ray gas opacities are the same in all three dust models. To compare $\zeta_\mathrm{X}$ to models without X-ray dust opacities they are also included in Fig.~\ref{fig:xionratesver}. 

As already mentioned, scattering of X-rays by dust can be neglected due to the strongly forward peaked scattering phase function (see Sect.~\ref{sec:dustopac}). As the dust acts only as an additional absorption agent, $\zeta_\mathrm{X}$ is reduced wherever the dust opacity becomes similar or larger than the gas opacity. This is in particular the case for high X-ray energies (see Fig.~\ref{fig:xdustopac}) but depends on the chosen dust properties. In the case of the small grains, dust absorption dominates the X-ray opacity for X-ray energies $E_\mathrm{X}\gtrsim 1~\mathrm{keV}$. As a consequence $\zeta_\mathrm{X}$ drops by about an order of magnitude and more, compared to the model with gas opacities only (see Fig.~\ref{fig:xionratesver}). For the medium and large grain models the impact of dust is much less severe. For the medium grains dust extinction becomes relevant for $E_\mathrm{X}\gtrsim 3~\mathrm{keV}$ and for the large grains only for $E_\mathrm{X}\gtrsim 10~\mathrm{keV}$, but in the large grains model, gas extinction remains always higher than dust absorption (see Fig.~\ref{fig:xdustopac}). 

Our results imply that dust extinction plays an important role for young disks, whereas for evolved disks with large grains, \mbox{X-rays} can penetrate deeper.
In evolved disks only the most energetic X-rays are affected as the gas opacity drops rapidly with energy. Therefore the impact on $\zeta_\mathrm{X}$ is the largest in the deep, high density, layers of the disk where $\zeta_\mathrm{X}$ is dominated by high energy X-rays. In young objects, where the disk is still embedded in an envelope, small grains can be important and should be included in X-ray radiative transfer models. We will investigate such a scenario based on the Class~I P{\tiny RO}D{\tiny I}M{\tiny O} model presented in \citet{Rab2017a} in a future study.
%
\subsubsection{Impact of X-ray background fields}
\label{sec:xionbg}
The importance of X-ray background fields for disks was estimated analytically by \citet{Adams2012}. They find that the X-ray background flux can be larger than the stellar X-ray flux for disk radii $r\gtrsim14\,\mathrm{au}$, assuming a geometrically flat disk and a typical disk impact angle for the stellar radiation. Their estimated radius corresponds to the radius where the stellar and interstellar X-ray flux becomes equal. However, attenuation by the disk itself was not taken into account (i.e. they compared the stellar and background X-ray fluxes at the disk surface).

\begin{figure}
\resizebox{\hsize}{!}{\includegraphics{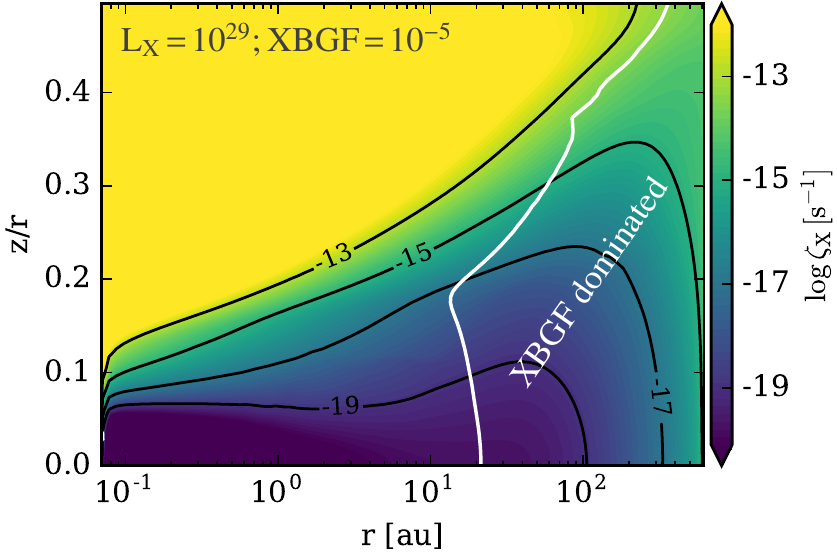}}
\caption{X-ray ionization rate $\zeta_\mathrm{X}$ for a model with $L_\mathrm{X}=10^{29}\mathrm{\,erg\,s^{-1}}$ and $F_\mathrm{XBGF}=2\times10^{-5}\mathrm{\,erg\,cm^{-2}s^{-1}}$ (i.e. the benchmark values of \citealt{Adams2012}). The white solid contour line encloses the region where the XBGF dominates $\zeta_\mathrm{X}$ (i.e. $\zeta_\mathrm{X,XBGF}\ge\zeta_\mathrm{X,*}$).}
\label{fig:BGzetaX}
\end{figure}
\begin{figure*}
\includegraphics{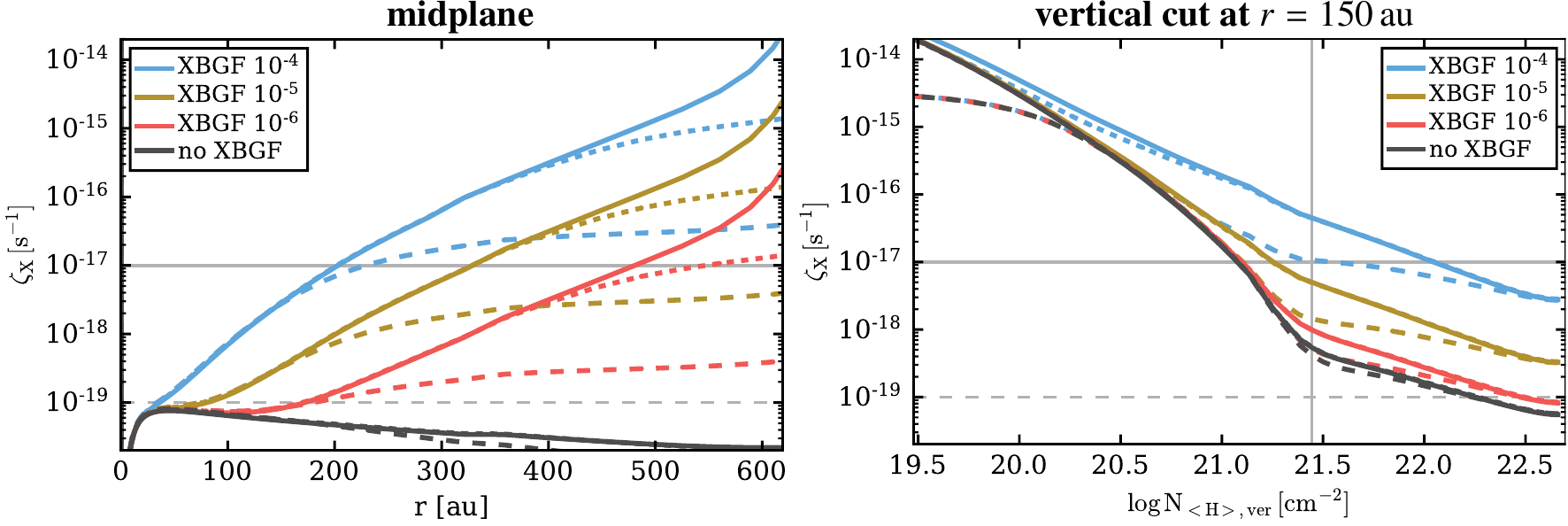}
\caption{X-ray ionization rate in the midplane (left panel) and for a vertical cut at $r=150\,\mathrm{au}$ (right panel) for models with fixed stellar X-ray luminosity ($L_\mathrm{X}=10^{30}\,\mathrm{erg\,s^{-1}}$) but varying X-ray background fields with fluxes of $2\times10^{-4}$ (blue), $2\times10^{-5}$ (brown) and $2\times10^{-4}$ $\mathrm{\,erg\,cm^{-2}s^{-1}}$ (red); the black lines are for the model without an XBGF. The solid lines are for the full X-ray spectra ($0.1-20\,\mathrm{keV}$); the dotted and dashed lines correspond to models with a low-energy cut-off at $0.3\,\mathrm{keV}$ and $\mathrm{1\,\mathrm{keV}}$, respectively. The horizontal lines mark the cosmic-ray ionization rates for the ISM ($\zeta_\mathrm{CR}\approx10^{-17}$), and low cosmic-ray case ($\zeta_\mathrm{CR}\approx10^{-19}\,\mathrm{s^{-1}}$). In the right panel the vertical grey solid line indicates the scattering surface at $N_\mathrm{\langle H\rangle,rad}=2\times10^{24}\,\mathrm{cm^{-2}}$ (see also Fig.~\ref{fig:xionratesver}).} 
\label{fig:zXBGmv}
\end{figure*}
In Fig.~\ref{fig:BGzetaX} we show $\zeta_\mathrm{X}$ for our full 2D disk model using $L_\mathrm{X}=10^{29}\,\mathrm{erg\,s^{-1}}$ and an X-ray background flux of $F_\mathrm{XBGF}=2\times10^{-5}\,\mathrm{erg\,cm^{-2}\,s^{-1}}$ (these are the values used by \citealt{Adams2012} for their benchmark case). We mark the region of the disk where the X-ray background field dominates $\zeta_\mathrm{X}$ (i.e. $\zeta_\mathrm{X,XBGF}\ge\zeta_\mathrm{X,*}$). In the midplane of the disk ($z=0$) the XBGF dominates for \mbox{$r\gtrsim20\,\mathrm{au}$}. Assuming a geometrically flat disk, the XBGF dominates in our model for $r\gtrsim15\,\mathrm{au}$ (we simply projected the smallest radius at the maximum height, where the XBGF dominates, to the midplane). For the same XBGF but $L_\mathrm{X}=10^{30}\,\mathrm{erg\,s^{-1}}$ we find that the XBGF dominates for $r\gtrsim30\,\mathrm{au}$ for a geometrically flat disk and \mbox{$r\gtrsim45\,\mathrm{au}$} for the midplane of our 2D model. Those radii are consistent with the analytical estimates of \citet{Adams2012}.

In Fig.~\ref{fig:zXBGmv} we show $\zeta_\mathrm{X}$ in the midplane and for a vertical cut at $r=150\,\mathrm{au}$ for models with different XBGF fluxes and a fixed stellar X-ray luminosity of $L_\mathrm{X}=10^{30}\,\mathrm{erg\,s^{-1}}$. Additionally we show models with a low energy cut-off for the stellar and XBGF spectrum at $0.3$ and $1\,\mathrm{keV}$, respectively. With this low energy cut-off we simulate (in a simple way) a possible absorption of the X-rays before they actually impinge on the disk. Such an absorption can happen by material close to the star (e.g. accretion columns \citealt{Grady2010b}) for the stellar X-rays and in case of the \mbox{X-ray} background field additional extinction due to the interstellar medium is also possible. The corresponding absorption columns required for those cut-offs are $N_\mathrm{\langle H\rangle}\approx10^{21}\,\mathrm{cm^{-2}}$ for $0.3\,\mathrm{keV}$ and $N_\mathrm{\langle H\rangle}\approx10^{22}\,\mathrm{cm^{-2}}$ for $1\,\mathrm{keV}$ (see Fig.~\ref{fig:xdustopac} and \citealt{Ercolano2009b}).

As seen from the left panel of Fig.~\ref{fig:zXBGmv}, $\zeta_\mathrm{X}$ at the outer radius of the disk can be as high as $10^{-14}\,\mathrm{s^{-1}}$ but strongly depends on the assumed low energy cut-off. \citet{Adams2012} also estimated $\zeta_\mathrm{X}$ for their benchmark X-ray background field ($F_\mathrm{XBGF}=2\times10^{-5}\,\mathrm{erg\,cm^{-2}\,s^{-1}}$) assuming an average X-ray photon energy of $E_\mathrm{X}=1\,\mathrm{keV}$, they find $\zeta_\mathrm{X}=8\times10^{-17}\,\mathrm{s^{-1}}$. This is similar to our model with the $0.3\,\mathrm{keV}$ low energy cut-off.

The low energy cut-off has no significant impact on the radius down to which the XBGF dominates $\zeta_\mathrm{X}$. For this high density regions only the most energetic X-rays, which are not affected by the low energy cut-off, are of relevance. However, as seen from Fig.~\ref{fig:zXBGmv} a possible absorption of the XBGF photons before they reach the disk has a significant impact for radii $r\gtrsim 200\,\mathrm{au}$ and at higher layers of the disk. For the $1\,\mathrm{keV}$-cut-off $\zeta_\mathrm{X}$ can be lower by more than an order of magnitude compared to the reference model with a minimum X-ray energy of $0.1\,\mathrm{keV}$.

More important than the value of $\zeta_\mathrm{X}$ at the outer disk radius is the value at higher densities where the actual emission from molecular ions originates. From Fig.~\ref{fig:zXBGmv} we can see that $\zeta_\mathrm{X}$ drops already below the ISM cosmic-ray ionization rate at $r\approx200\,\mathrm{au}$ even for the strongest XBGF considered. However, for the low cosmic-ray case the XBGF can be the dominant high energy ionization source in the midplane for radii as small as $r\approx50\,\mathrm{au}$. This is also seen in the right panel of Fig.~\ref{fig:zXBGmv}, depending on the XBGF flux $\zeta_\mathrm{X}$ can reach values around $5\times10^{-18}\,\mathrm{s^{-1}}$ close to the midplane of the disk at $r=150\,\mathrm{au}$. 
\subsection{Molecular ion column densities}
\label{sec:molcd}
In this Section we show results for the radial column density profiles of the disk ionization tracers HCO$^+$ and N$_2$H$^+$. We use these two molecules because they are the most commonly detected molecular ions in disks \citep[e.g.][]{Thi2004a,Dutrey2007a,Oberg2011a,Guilloteau2016} and because they trace different regions in the disk.

HCO$^+$ and N$_2$H$^+$ are mainly formed via proton exchange of $\mathrm{H_3^+}$ with CO and $\mathrm{N_2}$, respectively. The main destruction pathway is dissociative recombination with free electrons, where the metals (e.g. sulphur) play a significant role as additional electron donors \citep[e.g.][]{Graedel1982c,Teague2015b,Kamp2013b,Rab2017}. HCO$^+$ and N$_2$H$^+$ are sensitive to high energy ionization sources such as X-rays and cosmic-rays,  because the formation of $\mathrm{H_3^+}$ involves the ionization of $\mathrm{H_2}$ ($15.4\,\mathrm{eV}$ ionization potential).

Besides free electrons another efficient destruction pathway for N$_2$H$^+$ and $\mathrm{HCO^+}$ are ion-neutral reactions. N$_2$H$^+$ is efficiently destroyed by CO and therefore resides mainly in regions where gas phase CO is depleted (e.g. frozen-out). This makes it a good observational tracer of the CO ice line in disks \citep[e.g][]{Qi2013bc,Aikawa2015,vantHoff2017}. In case of $\mathrm{HCO^+}$, gas phase water is the destructive reaction partner. Observations of protostellar envelopes indicate that $\mathrm{HCO^+}$ is indeed sensitive to the water gas phase abundances \citep[e.g.][]{Jorgensen2013o,Bjerkeli2016,vanDishoeck2014}. In disks this is more difficult to observe, due to the more complex structure and because the water snow line in disks is located at much smaller radii ($r\approx1\,\mathrm{au}$ for a T Tauri star) compared to CO ($r\approx20\,\mathrm{au}$). However, $\mathrm{HCO^+}$ follows mainly the distribution of gas phase CO in the disks, whereas N$_2$H$^+$ traces regions where CO is frozen-out (i.e. where the temperature is $\lesssim25\,\mathrm{K}$).

\begin{figure*}
\centering
\includegraphics{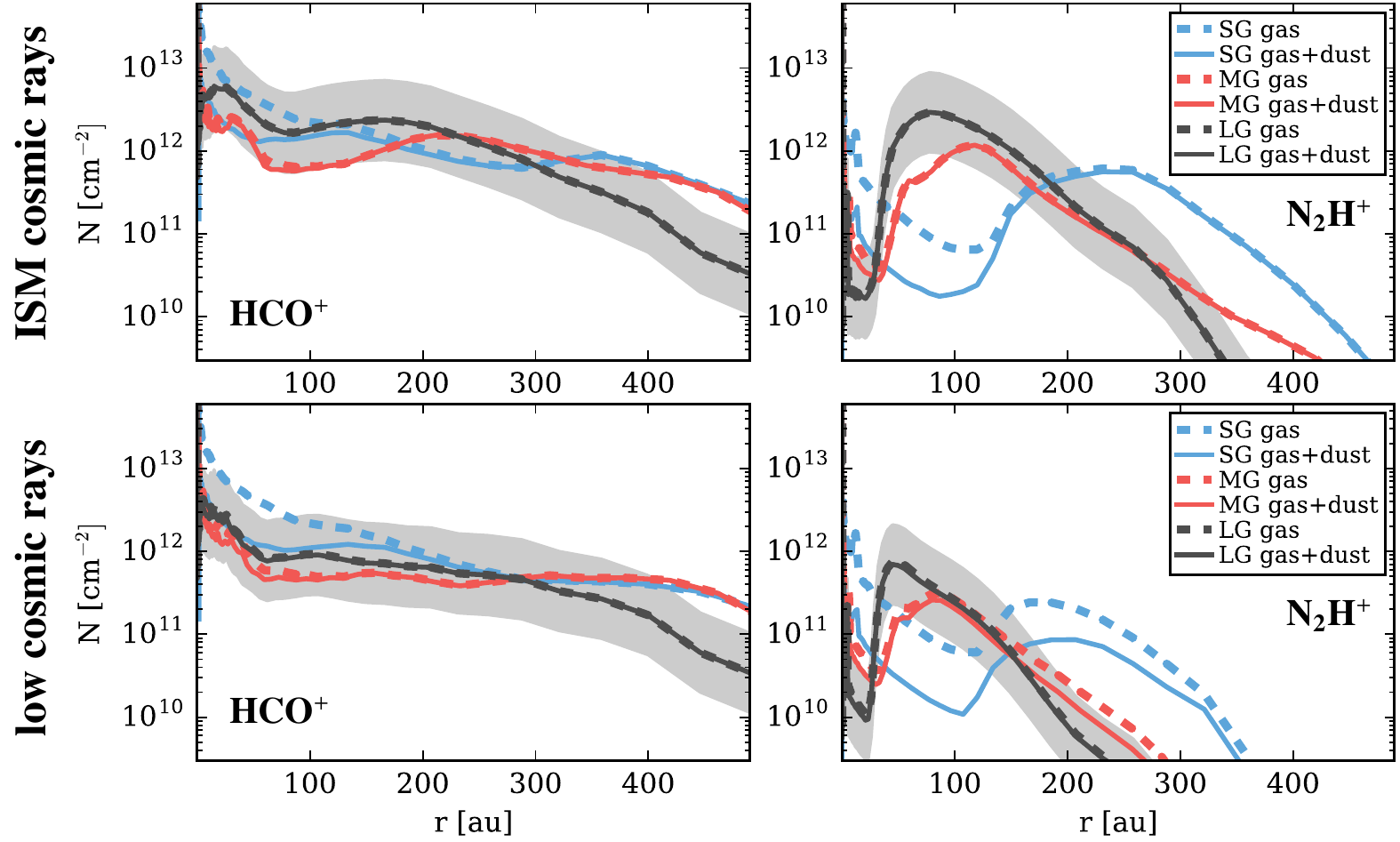}
\caption{HCO$^+$ (left column) and N$_2$H$^+$ (right column) radial column density profiles for models with different dust size distributions: small grains (SG, blue), medium grains (MG, red) and large grains (LG, black). The dashed lines are for models where only the gas component is considered in the X-ray RT, where the solid lines show models where both X-ray gas and dust opacities are included. The ISM cosmic-ray models are shown in the top row, the low cosmic-ray models in the bottom row. The grey shaded area marks a difference of a factor three in $N$ with respect to the reference model (LG gas+dust).}
\label{fig:mcd_dust}
\end{figure*}
For the abundance of the molecular ions also the so called sink-effect for CO and $\mathrm{N_2}$ is of relevance. The main mechanism of the sink-effect is the conversion of CO and N$_2$ to less volatile species which freeze-out at higher temperatures or remain on the dust grains. This can happen via surface chemistry and/or via dissociation of neutral molecules by He$^+$ \citep[e.g.][]{Aikawa1996h,Bergin2014,Cleeves2015a,Helling2014,Furuya2014g,Reboussin2015b,Aikawa2015}. The main consequence of the sink-effect is the depletion of gas phase CO and N$_2$ in regions with temperatures above their respective sublimation temperatures. However, the efficiency of the sink-effect is not very well understood as it depends on various chemical parameters (see \citealt{Aikawa2015} for more details). In our model only the He$^+$ sink-effect is considered.

Our model also includes excited $\mathrm{H_2}$ chemistry that opens up another formation pathway for $\mathrm{HCO^+}$. This formation pathway can be important close to the $\mathrm{C^+/C/CO}$ transition (see \citealt{Greenwood2017} and Appenix~\ref{sec:HCOpHtexc}). The relevance of this pathway will be discussed later on.

The typical abundance structure for HCO$^+$ and N$_2$H$^+$ in our reference model is presented in \citet{Rab2017}. Here we focus on the radial column densities because they can be more easily compared to observations and other thermo-chemical disk models. 
\subsubsection{Impact of dust grain size distributions on chemistry}
\label{sec:mcd_dust}
In Fig.~\ref{fig:mcd_dust} we show the molecular ion column densities $N_\mathrm{HCO^+}$ and $N_\mathrm{N_2H^+}$ for the three different dust models, small grains (SG), medium grains (MG) and large grains (LG) described in Sect.~\ref{sec:dustmodels}. For each dust model also both cases of cosmic-ray ionization rates, low and ISM cosmic rays are shown (Sect.~\ref{sec:crmodels}). Further we show models with and without \mbox{X-ray} dust opacities. All models shown in Fig.~\ref{fig:mcd_dust} have \mbox{$L_\mathrm{X}=10^{30}\,\mathrm{erg\,s^{-1}}$} and no X-ray background field.

It is clearly seen in Fig.~\ref{fig:mcd_dust} that neither for the medium grains nor for large grains the inclusion of X-ray dust opacities has a significant impact on the molecular ion column densities. Only for $\mathrm{N_2H^+}$ a slight decrease on the column density can be seen in the models with low cosmic rays (e.g. compare model MG gas with MG gas+dust). The column densities are not significantly affected by including X-ray dust opacities  as the strongest impact of the dust on $\zeta_\mathrm{X}$ is limited to regions close to the midplane (see Fig.~\ref{fig:xionratesver}). There CRs mostly dominate the molecular ion abundances as the X-ray ionization is typically $\zeta_\mathrm{X}\lesssim10^{-19}\,\mathrm{s^{-1}}$, even if dust opacities are not included. Further the contribution to the molecular ion column densities from regions close to the midplane is limited as the parent molecules of the ions are frozen out anyway. The situation is different for the SG model, where $\zeta_\mathrm{X}$ is affected by the dust at all disk layers (see Fig.~\ref{fig:xionratesver}) and $N_\mathrm{HCO^+}$ and $N_\mathrm{N_2H^+}$ can drop by factors of three to ten. This shows that it is justified, in case of HCO$^+$ and N$_2$H$^+$, to neglect X-ray dust opacities for evolved disk dust populations but not necessarily for ISM like dust.

Fig.~\ref{fig:mcd_dust} also shows that, independent of the X-ray dust opacities, the dust grain size distributions themselves have a significant impact on the molecular column densities. In the SG model the gas disk is more efficiently shielded from the stellar and interstellar UV radiation field but also the total dust surface per hydrogen nucleus increases significantly (see Table~\ref{table:models}). This has mainly two consequences: 

Firstly, the ionization of metals such as carbon and sulphur is significantly reduced. This causes a decrease in the number of free electrons available for the dissociative recombination with molecular ions. On the other hand the impact of the dust opacities on the X-ray disk radiation field (and on $\zeta_\mathrm{X}$) is less significant (SG model), or not significant at all (MG and LG model) compared to the impact of the dust on the UV radiation field. Consequently the abundance of the molecular ions increases in regions which are efficiently shielded from the UV radiation fields by the presence of small grains.

Secondly, the freeze-out and the sink-effect become more important if the total dust surface increases. This reduces the abundance of molecular ions in high density regions that are efficiently shielded from UV radiation (i.e. no photo-desorption).

These effects are best seen for $N_\mathrm{N_2H^+}$. Compared to the LG model, in the SG model the abundance of $\mathrm{N_2H^+}$ is reduced close to the midplane of the disk due to the sink-effect and freeze-out (i.e. lower gas phase abundance of the parent molecule N$_2$) but increases in the outer and upper layers of the disk due to the shielding of the UV radiation by small grains (i.e. lower abundance of metal ions). This results in a shift of the $N_\mathrm{N_2H^+}$ peak to larger radii ($r\approx150-200\,\mathrm{au}$), and the peak is not tracing the radial CO ice line anymore (which is at $r\approx45\,\mathrm{au}$ in the SG model). For radii $r\lesssim150\,\mathrm{au}$, $N_\mathrm{N_2H^+}$ is now dominated by the $\mathrm{N_2H^+}$ layer just below the C$^+$/C/CO transition where the X-ray ionization rate is high enough so that $\mathrm{N_2H^+}$ survives also in layers with gas phase CO (see \citealt{Aikawa2015,vantHoff2017,Rab2017}). In contrast to the LG  grain model, $N_\mathrm{N_2H^+}$ in the SG model can reach comparable or even higher values in the inner disk ($r\lesssim100\,\mathrm{au}$) compared to the peak value around $r\approx150-200\,\mathrm{au}$. 

\citet{Aikawa2015} also used two different dust size distributions (ISM like and large grains) for their detailed study on $\mathrm{N_2H^+}$ in protoplanetary disks. Their resulting column density profiles are very similar to what is shown in Fig.~\ref{fig:mcd_dust}. In the the chemical models of \citet{Dutrey2007a} for \object{DM Tau} and \object{LkCa 15} the peak in their $N_\mathrm{N_2H^+}$ profiles are at very large radii ($r\gtrsim400\,\mathrm{au}$), which is likely due to their assumed single grain size of $0.1\,\mathrm{\mu m}$. \citet{Cleeves2015a} used a reduced dust surface area, compared to $0.1\,\mathrm{\mu m}$ grains, to model $N_\mathrm{N_2H^+}$ for \object{TW Hya}, however they also required a  lower cosmic-ray ionization rate ($\zeta_\mathrm{CR}\approx10^{-19}\,\mathrm{s^{-1}}$) to match the observed sharp peak in $N_\mathrm{N_2H^+}$, located close to the CO ice line. A lower cosmic-ray ionization rate decreases the efficiency of the He$^+$ sink-effect. A strong impact of the low-cosmic ray ionization rate on the $N_\mathrm{N_2H^+}$ peak is not really seen in our models or in the models of \citet{Aikawa2015}. This might be caused by differences in the time-scales for the sink-effect \citep{Bergin2014}. In our LG model, steady-state for $N_\mathrm{N_2H^+}$ is already reached at a chemical age of approximately $1\,\mathrm{Myr}$ \citep[see][]{Rab2017}. For the SG models a time-dependent test run with ISM cosmic rays showed that steady-state is only reached after $2-3\,\mathrm{Myr}$ in regions around the radial CO ice line, whereas at $\approx1\,\mathrm{Myr}$ the $N_\mathrm{N_2H^+}$ peak is still tracing the radial CO ice line.

Despite the differences in the various chemical models, they all indicate that the sink-effect plays a crucial role for the shape of the $\mathrm{N_2H^+}$ radial column density profile. Further, only models accounting for dust growth are able to reproduce a sharp peak in the $N_\mathrm{N_2H^+}$ profile  near the radial CO ice line as is observed for \object{TW Hya} (\citealt{Qi2013bc}, but see also \citealt{Aikawa2015,vantHoff2017} for a discussion on the robustness of $N_\mathrm{N_2H^+}$ as a CO ice line tracer). This is consistent with dust observations clearly indicating grain growth and dust settling in disks \citep[e.g.][]{Andrews2005q,Pinte2016}. In any case the chemical modelling results for $N_\mathrm{N_2H^+}$ indicate that $\mathrm{N_2H^+}$ is not only a tracer of the radial CO ice line but also for dust evolution in disks.
\begin{figure*}
\centering
\includegraphics{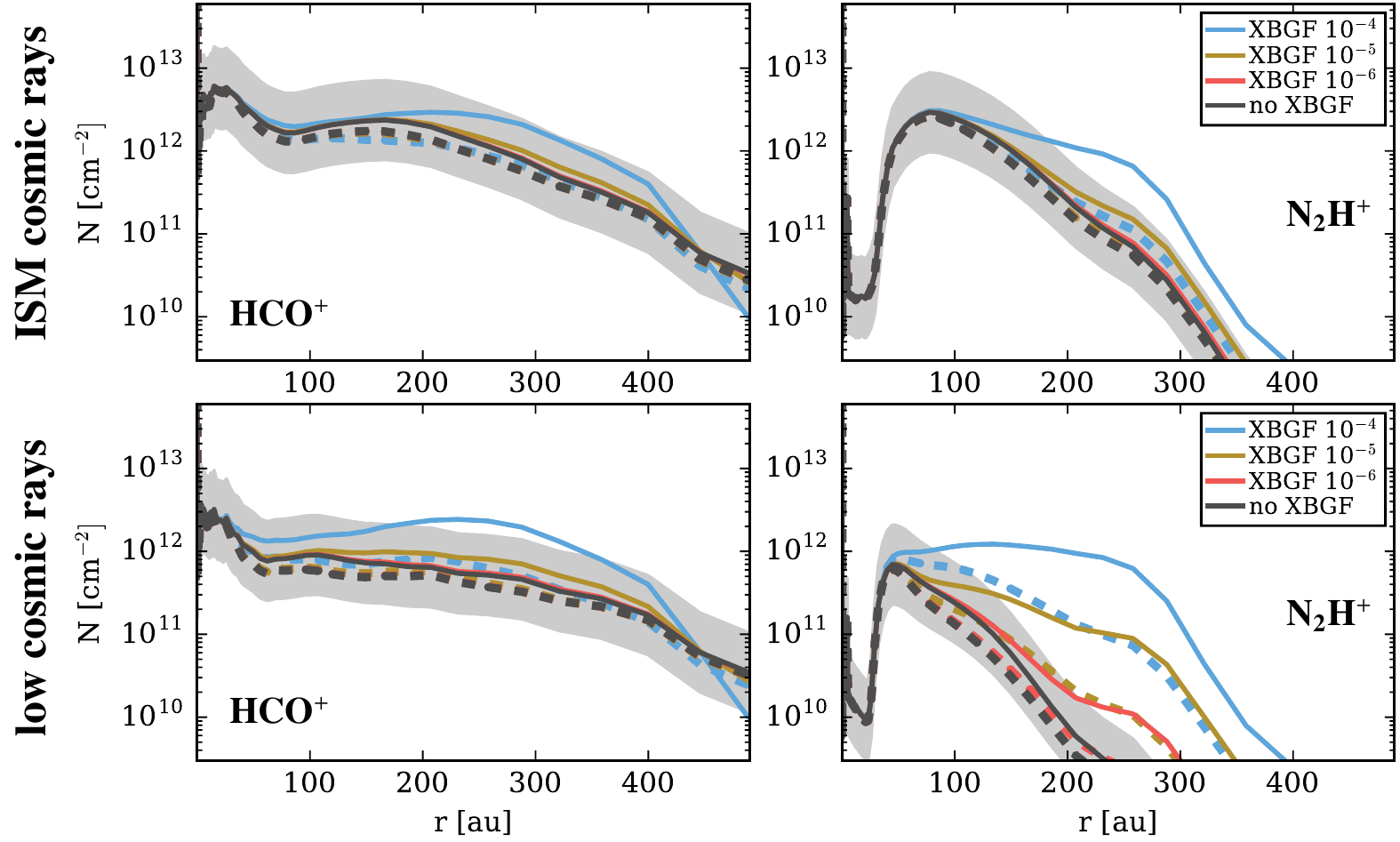}
\caption{Radial column density profiles for HCO$^+$ and $\mathrm{N_2H^+}$ with (coloured lines) and without (black lines) XBGF for the ISM (top row) and low (bottom row) cosmic-ray case. The dashed lines are for models with a low-energy cut-off of the X-ray spectra (stellar and XBGF) of $1\,\mathrm{keV}$, for the solid lines the lowest X-ray energy is $0.1\,\mathrm{keV}$. The grey shaded area marks a factor of three difference in $N$ with respect to the models without XBGF (the solid black line).}
\label{fig:molXBGF}
\end{figure*}
\subsubsection{Impact of X-ray background fields on chemistry}
\label{sec:XBGFresults}
To show the impact of X-ray background fields (XBGF) on the molecular column densities we compare in Fig.~\ref{fig:molXBGF} models with $L_\mathrm{X}=10^{30}\,\mathrm{erg\,s^{-1}}$ but varying XBGF fluxes. We also include models with a low-energy cut-off at $1\,\mathrm{keV}$ to show the impact of the possible absorption of soft X-rays before they reach the disk (see also Sect.~\ref{sec:xionbg}). For each of these models the results for low and ISM cosmic rays are shown. For all models the large grains dust model is used.

For the case of the ISM cosmic-ray ionization rate the impact of the XBGF on the column densities is limited. Only for models with the highest XBGF flux of $2\times10^{-4}\,\mathrm{erg\,cm^{-2}\,s^{-1}}$ the column densities increase by more than a factor of three for radii $r\gtrsim250\,\mathrm{au}$. Although the XBGF dominates the X-ray ionization rate down to $r\approx\,20\,\mathrm{au}$, $\zeta_\mathrm{X}>\zeta_\mathrm{CR}$ is true only for $r\gtrsim200\,\mathrm{au}$, and only for the case of the strongest XBGF (see Fig.~\ref{fig:zXBGmv}).

The impact of the XBGF is much larger in the case of a low CR ionization rate. In that case the molecular ion column densities are generally lower compared to ISM CRs and the relative impact of the XBGF increases. However, the impact on $N_\mathrm{HCO^+}$ remains limited; only for the strongest XBGF, $N_\mathrm{HCO^+}$ increases by about a factor of five at most. For $N_\mathrm{N_2H^+}$ the picture is quite different. Due to the low CRs the column densities for $r\gtrsim200\,\mathrm{au}$ are reduced by more than an order of magnitude compared  to the ISM CRs models. In these regions the XBGF is now most effective and consequently $N_\mathrm{N_2H^+}$ increases significantly. For the strongest XBGF $N_\mathrm{N_2H^+}$ increases by up to two orders of magnitude for $r\gtrsim200\,\mathrm{au}$ and reaches levels similar to the ISM CR models. The reason why $\mathrm{N_2H^+}$ is more sensitive to the high energy ionization sources is its location in the disk. Compared to HCO$^+$, $\mathrm{N_2H^+}$ is mainly located in deeper layers of the disk; below the CO ice line. In those layers, the ionization balance is mostly dominated by molecular ions, as the ionization of atomic metals by UV becomes less important.

In the models with a low-energy cut-off at $1\,\mathrm{keV}$ for the X-ray spectra, HCO$^+$ is not affected by the XBGF even in the low CR model. Also the impact on $N_\mathrm{N_2H^+}$ is now weaker. $N_\mathrm{N_2H^+}$ is typically a factor of a few up to an order of magnitude lower compared to the models with a cut-off at $0.1\,\mathrm{keV}$. Although we use the low-energy cut-off also for the stellar X-rays, such a drop in the column densities is not seen in the models without XBGFs. The reasons are the geometrical dilution of the stellar X-ray radiation and that the stellar X-rays have to penetrate the high radial and vertical column densities in the inner disk. The XBGF irradiates the disk isotropically and only has to penetrate the low column densities of the outer disk. Therefore also the low-energy X-rays can penetrate larger areas of the disk and have more impact on $\zeta_\mathrm{X}$ in disk regions relevant for the molecular ions. 

For HCO$^+$ we actually also see a drop in the column density for $r\gtrsim400\,\mathrm{au}$ for the strongest XBGF. The reason for this is a lower CO abundance caused by X-ray photo-dissociation which is also included in our chemistry model. The CO abundance at $r\gtrsim450\,\mathrm{au}$ drops by factors of approximately three to five down to heights of $z\approx20\,\mathrm{au}$, which results in a drop of the HCO$^+$ abundance by nearly an order of magnitude. The situation is similar for N$_2$ and $\mathrm{N_2H^+}$, however the abundance of $\mathrm{N_2H^+}$ is already below $10^{-12}$.

As noted (Sect.\ref{sec:molcd}) our model includes also chemistry of excited molecular hydrogen $\mathrm{H_2^*}$. This opens up a formation channel for HCO$^+$ via the ion-neutral reaction of $\mathrm{H_2^*}$ with $\mathrm{C^+}$ (see Appendix~\ref{sec:HCOpHtexc} for details). In the inner disk this reaction is only effective in a very thin layer at the C$^+$/C/CO transition. However, in the outer disk for $r\gtrsim300\,\mathrm{au}$ this reaction becomes significant, as the C$^+$/C/CO transition is not as sharp as in the inner disk. If this reaction is deactivated, we find that for such a model the slope of $N_\mathrm{HCO^+}$ for $r>300\,\mathrm{au}$ becomes steeper. However, our findings concerning the impact of XBGFs are not significantly affected, $N_\mathrm{HCO^+}$ becomes slightly more sensitive to XBGFs for $r\gtrsim300\,\mathrm{au}$ and X-ray photo-dissociation becomes less significant for $N_\mathrm{HCO^+}$ at large radii where $N_\mathrm{HCO^+}$ drops below $10^{10}\,\mathrm{cm^{-2}}$ ($r\approx400\,\mathrm{au}$). The results for $\mathrm{N_2H^+}$ are not affected by $\mathrm{H_2^*}$ chemistry.
\section{Discussion}
\label{sec:discussion}
\subsection{Observational implications of X-ray background fields}
Our results indicate that the molecular ion column densities in the outer disk are sensitive to X-ray background fields. At least for $\mathrm{N_2H^+}$ the effect on the column density can be strong enough to be observable with modern (sub)millimetre telescopes like ALMA (Atacama Large Millimeter Array), NOEMA (NOrthern Extended Millimeter Array) and SMA (Submillimeter Array). However, our results also show that certain conditions such as a strong XBGF and low CR ionization rates are required.

The typical value for the X-ray background flux in young embedded clusters of $2\times10^{-5}\,\mathrm{erg\,cm^{-2}\,s^{-1}}$ corresponds to cluster member sizes with $N<3000$, typical for clusters in the solar vicinity \citep{Adams2012}. However, this value drops to $\lesssim10^{-5}\,\mathrm{erg\,cm^{-2}\,s^{-1}}$ for smaller clusters with $N\approx100$, which is more typical for the Taurus star formation region;  the X-ray XEST survey detected 136 X-ray sources \citep{Gudel2007d}. A typical disk in Taurus would only be affected by an XBGF if it is shielded efficiently from cosmic-rays (see Fig.~\ref{fig:molXBGF}). However, even in a small cluster some of the sources might experience a higher X-ray background flux depending on their location within the cluster (e.g. closer to the cluster center, see \citealt{Adams2012}). To identify such sources a large sample of spatially resolved $\mathrm{N_2H^+}$ observations (see below) would be required. The current number of $\mathrm{N_2H^+}$ detections in protoplanetary disk is less than ten \citep{Dutrey2014} and it is therefore not surprising that signatures of XBGF have not been found yet. 

Due to the rather poor knowledge concerning the impact of individual high-energy ionization sources on disk chemistry (see e.g. \citealt{Rab2017}) and as can only the stellar X-ray luminosity can be measured, it will be certainly challenging to discriminate the influence of XBGFs from the stellar high-energy ionization sources (X-rays and stellar energetic particles), cosmic-rays and radionuclide ionization. However, all these ionization sources are different in the way they irradiate the disk.

The stellar high energy ionization sources have less impact on the outer disk; they need to penetrate the high densities in the inner disk and also experience geometric dilution as they irradiate the disk as a point source (see also \citealt{Rab2017}). Cosmic rays irradiate the disk isotropically and therefore have an impact on the whole disk. Due to their high energies, CRs are only significantly absorbed in the inner disk ($r\lesssim10\,\mathrm{au}$) and provide therefore a nearly constant $\mathrm{H_2}$ ionization rate for the bulk of the disk material (this is similar for radionuclide ionization). In contrast to CRs, the XBGF experiences significant absorption by the disk material even in the outer disk. Those differences result in different shapes of the $\mathrm{N_2H^+}$ radial column density profiles as is seen in Fig.~\ref{fig:molXBGF}. XBGFs make the profile shallower as the impact on the molecular column densities decreases for smaller disk radii (measured from the star). In contrast, a change in the CR ionization rate affects the radial column density profile at all radii. 

In any case modelling of spatially resolved $\mathrm{N_2H^+}$ observations of disks are required to discriminate the contribution of XBGF to disk ionization. Such observations for T Tauri disks are still rare \citep{Dutrey2007a,Qi2013bc} but certainly will become more common in the near future due to the significant advances, in particular in terms of spatial resolution, of modern (sub)mm interferometers.%
\subsection{UV background fields and external photo-evaporation}
For all models presented here we assumed the canonical value for the interstellar UV radiation of $\chi^\mathrm{ISM}=1$ (in units of the Draine field). This is a reasonable assumption for low-mass star formation regions as an enhanced UV background field would be mostly produced by massive stars with spectral type O, B and A. In contrast to the UV field, the X-ray background field is mainly produced by X-ray emission of solar-like low mass stars (see \citealt{Adams2012}). However, a presence of an enhanced UV background field certainly has an impact on the outer disk due to photo-dissociation of molecules, photo-desorption of ices and photo-ionization of metals (see e.g. \citealt{Teague2015b}). These processes will reduce the abundances of molecular ions in the outer disk as their parent molecules are dissociated and metals like carbon and sulphur are ionized and will therefore dominate the ionization balance. Consequently, the presence of an enhanced UV background field will lower the impact of the XBGF on the outer disk molecular abundances. In high-mass star formation regions like Orion with UV background fields up to $\chi^\mathrm{ISM}\approx10^4$, most likely the UV field will completely dominate the chemistry in the outer disk \citep{Walsh2013a,Antonellini2015}.

There is evidence that already weak UV background fields $\chi^\mathrm{ISM}\gtrsim4$ can drive photo-evaporation of the outer disk (\mbox{\citealt{Haworth2017}}, but see also \citealt{Adams2010a,Williams2011}). Such a process is not included in our model. However, we do not expect a huge impact of photo-evaporation on our results. For the X-ray RT mainly the column density matters that photons have to penetrate until they reach the regions of the disk where the molecular ions become abundant. These column densities are not significantly affected if parts of the outer disk material would slowly drift radially outwards.

On the other hand XBGFs can act as an additional heating source for the outer disk and therefore could contribute to external disk photo-evaporation. We find that only the strongest X-ray background field with $F_\mathrm{XBGF}=2\times10^{-4}\,\mathrm{erg\,cm^{-2}\,s^{-1}}$ has a significant impact on the gas temperature in our models. In the midplane of the disk at $r\approx500\,\mathrm{au}$ ($n_\mathrm{\langle H\rangle}\approx4\times10^5\,\mathrm{cm^{-3}}$), the gas temperature increases from about $T_\mathrm{g}\approx24\,\mathrm{K}$ to $T_\mathrm{g}\approx36\,\mathrm{K}$, where for the canonical XBGF we find only an increase of about $2\,\mathrm{K}$ (always $\chi^\mathrm{ISM}=1$ is used). Those gas temperatures are lower than the typical temperatures reported by \citet{Facchini2016} derived from hydrodynamical disk photo-evaporation models with $\chi^\mathrm{ISM}\geq30$. This indicates that in the presence of a modestly enhanced UV background field, XBGFs would play a rather minor role for external disk photo-evaporation. However, a more detailed investigation of the possible importance of XBGFs for external disk photo-evaporation is required to be more quantitative.
\subsection{The role of dust evolution}
\label{sec:dustevol}
The properties of the dust and its evolution has a significant impact on the chemical structure of the disk \citep[e.g.][]{Vasyunin2011,Akimkin2013,Woitke2016}. Important dust evolution processes are dust growth, dust settling and radial migration of (sub)mm sized grains. Individual disks likely will show different dust properties as dust evolution proceeds over time and also depends on various parameters such as the turbulence in the disk (see \citealt{Birnstiel2016} for a recent review). It is, therefore, crucial to have good knowledge of the dust disk (e.g. from continuum observations) for modelling of molecular line emission of individual protoplanetary disks. In the following, we briefly discuss the impact of dust evolution on our results and possible shortcomings of our dust model.

In our modelling approach, we do include methods to simulate dust growth and dust settling (see Sections \ref{sec:dustdisk} and \ref{sec:dustmodels}). The models with varying grain size distributions indicate a significant impact of the assumed grain size on the radial column densities of $\mathrm{HCO^{+}}$ and $\mathrm{N_2H^+}$ (see Sect.~\ref{sec:mcd_dust}). We also did run disk models using the large grains dust model but varied the degree of settling (i.e. no settling at all and strong settling with $\alpha_\mathrm{settle}=10^{-4}$). For the case of strong settling, the $\mathrm{HCO^{+}}$ and $\mathrm{N_2H^+}$ column densities change at most by a factor of three where $\mathrm{HCO^{+}}$ is more sensitive than $\mathrm{N_2H^+}$. In the non-settled disk model, only $\mathrm{N_2H^+}$ is affected where the column density drops by about an order of magnitude around $r=100\,\mathrm{au}$. However, the relative impact of X-ray background fields on the molecular column densities is similar to the models presented in Sect.~\ref{sec:XBGFresults} and consequently our conclusions concerning the impact of X-ray background fields are not affected. Nevertheless, for modelling of molecular observations of individual disks, it is also necessary to consistently model the dust component. The model presented here is well suited for such modelling, as the same dust model is used for the radiative transfer (X-ray to mm), the chemistry and also for the calculation of synthetic observables such as spectral energy distributions and spectral line emission (see \citealt{Woitke2016}).

Not included in our model is the inward radial migration of large dust grains. Spatially resolved (sub)mm observations indicate that the radial extent of dust emission is significantly smaller than the molecular gas emission \citep[for CO see e.g.][]{deGregorio-Monsalvo2013a,Cleeves2016}. However, models indicate that the difference in the extent of the dust and gas emitting areas might be an optical depth effect \citep{Woitke2016,Facchini2017}. Nevertheless, radial migration of large dust particles can have an impact on the molecular abundances in the outer disk as shown by \citet{Cleeves2016b} for CO. The models of \citet{Cleeves2016b} show that the depletion of large dust grains in the outer disk can increase the CO gas phase column density by about a factor of two, compared to models without radial migration. This will likely also have an impact on the $\mathrm{HCO^{+}}$ and $\mathrm{N_2H^+}$ column densities.

We do not expect that radial migration affects our general conclusions on the impact of X-ray background fields. The changes in the molecular ion abundances in the outer disk are purely driven by an increase of the X-ray ionization rate. It is unlikely that the X-ray ionization rate is strongly affected by the depletion of large dust grains in the outer disk, as in that case the X-ray opacities are dominated by the gas, similar to our large dust grains model (see Fig.~\ref{fig:xdustopac}). The impact of radial migration on the molecular ion abundances is an interesting aspect which certainly deserves more detailed investigations. However, such a study is out of the scope of this paper.

Recent ALMA observations have shown azimuthally symmetric gap- and ring structures in the disk emission of (sub)mm sized dust grains \citep[e.g.][]{ALMAPartnership2015,Andrews2016,Isella2016}. Although the dust depletion in the gaps has an impact on the molecular gas phase abundances \citep[e.g.][]{Teague2017}, those rings are rather narrow (width $\lesssim\,10\mathrm{au}$, \citealt{Pinte2016}) and the global radial distribution of the molecular column densities remains intact. Therefore our main conclusions concerning the impact of \mbox{X-ray} background fields are not significantly affected by the presence of gaps.
\section{Summary and conclusions}
\label{sec:conclusions}
We introduced a new X-ray radiative transfer module for the radiation thermo-chemical disk code P{\tiny RO}D{\tiny I}M{\tiny O}. This new module includes X-ray scattering and a detailed treatment of X-ray dust opacities, which can be applied to different dust compositions and dust size distributions. We investigated the importance of X-ray scattering, X-ray dust opacities and X-ray background fields of embedded young clusters for the X-ray ionization rates by means of a representative \mbox{T Tauri} protoplanetary disk model. Further, we studied the impact of X-ray background fields on the disk chemistry, where we used the observed disk ionization tracers HCO$^+$ and N$_2$H$^+$. Our main conclusions are:
\begin{itemize}
  \item For evolved dust size distributions (e.g. including grain growth) X-ray opacities are mostly dominated by the gas, only for energies $E_\mathrm{X}\gtrsim5-10\,\mathrm{keV}$ dust opacities become relevant. Consequently the disk X-ray ionization rates are only reduced in high density regions  close to the midplane if dust opacities are included in the X-ray RT. For ISM like dust size distributions (i.e. maximum grain size $\lesssim0.1\,\mathrm{\mu m}$) the X-ray ionization rate is affected throughout the disk and is reduced by more than an order of magnitude due to efficient absorption of X-rays by the dust.
  \item For a typical X-ray background flux, as expected for embedded young clusters in the solar vicinity, the XBGF dominates the X-ray ionization rate in the disk down to $r\approx20\,\mathrm{au}$. This is consistent with the analytical estimates of \citet{Adams2012}. However, due to absorption by the disk material, the XBGF ionization rate drops already below the ISM cosmic-ray ionization rate ($\zeta_\mathrm{CR} \approx10^{-17}\,\mathrm{s^{-1}}$) at $r\approx200-300\,\mathrm{au}$.
  \item XBGFs can have a significant impact on the vertical column densities of HCO$^+$ and N$_2$H$^+$, where N$_2$H$^+$ is most sensitive. However, this is only the case for a low cosmic-ray ionization rate $\zeta_\mathrm{CR} \approx10^{-19}\,\mathrm{s^{-1}}$, or for XBGF fluxes at least ten times stronger than the typical value. In case of an enhanced UV background field, the outer disk molecular chemistry would be strongly affected due to photo-dissociation and photo-ionization and the impact of XBGFs might become insignificant.
  \item Our results indicate that for an ordinary disk in a typical low-mass star formation region like Taurus the expected \mbox{X-ray} background flux has likely only little impact on the disk chemistry. However, it is possible that a fraction of the cluster members experience stronger fluxes depending on their location within the cluster.
  \item The various high energy ionization sources relevant for disk ionization are different in their energetic properties (e.g. \mbox{X-rays} versus cosmic-rays) and in the way they irradiate the disk (e.g. point source versus isotropic irradiation). This makes it possible to discriminate the individual contributions of the various sources, including X-ray background fields, to disk ionization from observations. Besides the modelling of spatially resolved molecular ion emission, also good knowledge of the stellar and disk properties (i.e. dust population) of the individual targets is required to achieve this.
\end{itemize}
With P{\tiny RO}D{\tiny I}M{\tiny O} it is now possible to model several high-energy disk ionization processes including cosmic rays, radionuclide ionization (not presented in this study), stellar energetic particles \citep{Rab2017} and X-rays (stellar and interstellar). However, further observational constraints are required to constrain the disk ionization fraction and to better understand molecular ion chemistry. Next to a good knowledge of the stellar (i.e. UV and X-ray radiation) and disk properties, spatially resolved observations of molecular ions such as N$_2$H$^+$ and HCO$^+$ are most crucial. Such observations are still rare but will likely be quite common in the near future due to the modern (sub)millimetre interferometers like ALMA, NOEMA and SMA.
\begin{acknowledgements} The authors thank the anonymous referee for useful and constructive comments which improved the paper. We want to thank A. Tielens and J. Slavin for providing the hot gas spectrum shown in Fig.~\ref{fig:BgSpectrum}. The research leading to these results has received funding from the European Union Seventh Framework Programme FP7-2011 under grant agreement no 284405. CHR and MG acknowledge funding by the Austrian Science Fund (FWF): project number P24790. The computational results presented have been achieved using the Vienna Scientific Cluster (VSC). This publication was supported by the Austrian Science Fund (FWF). This research has made use of NASA's Astrophysics Data System. All figures were made with the free Python module Matplotlib \citep{Hunter2007}. This research made use of Astropy, a community-developed core Python package for Astronomy \citep{AstropyCollaboration2013}.
\end{acknowledgements}

\bibliographystyle{aa}
\bibliography{xrt}
\begin{appendix}

\section{Comparison of X-ray photoelectric cross-sections}
\label{sec:xcscomp}
\begin{figure}
\resizebox{\hsize}{!}{\includegraphics{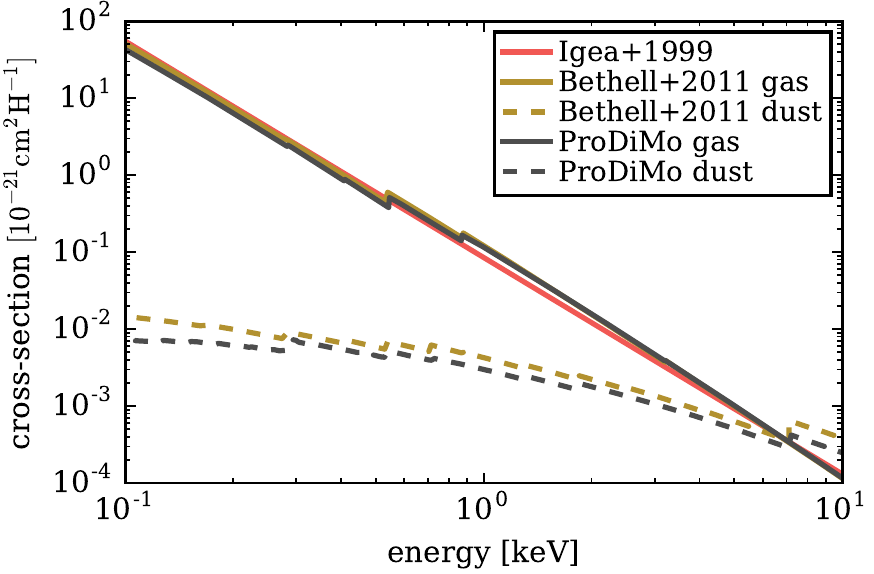}}
\caption{Comparison of X-ray photoelectric (absorption) cross-sections per hydrogen nucleus as a function of photon energy. The red line shows the gas absorption cross-section from \citet{Igea1999}. The brown lines show the cross-sections derived by \citet{Bethell2011a}. The black lines show the P{\tiny RO}D{\tiny I}M{\tiny O} results for the large grains dust model.}
\label{fig:xcscomp}
\end{figure}
In Fig.~\ref{fig:xcscomp}, we compare the X-ray photoelectric cross-sections as used in P{\tiny RO}D{\tiny I}M{\tiny O} to the cross-sections derived by \citet{Igea1999} and \citet{Bethell2011a}. For the P{\tiny RO}D{\tiny I}M{\tiny O} case we used the depleted element abundances \citep{Kamp2017} assuming all hydrogen is in $\mathrm{H_2}$ and all other elements are present as neutral atoms. For the dust, we use the large grains dust model (see Table~\ref{table:discmodel}) and the opacity calculations as described in Sect.~\ref{sec:dustopac}. 

For the \citet{Igea1999} cross-sections, we used their Eq.~10 for the case of heavy element depletion. \citet{Igea1999} did not include a treatment of the absorption edges and also neglected the dust component. For the \citet{Bethell2011a} cross-sections we used their Eq.~1 with the corresponding fitting coefficients for their ``Gas'' case (their Table~2); for the dust we used the coefficients for their dust size distribution with $a_\mathrm{min}=0.01\,\mathrm{\mu m}, a_\mathrm{max}=1000\,\mathrm{\mu m}$ and $a_\mathrm{pow}=3.5$  (their Table~3). We note that \citet{Bethell2011a} did not use  Mie-theory for the opacity calculations bur rather used directly the photo-electric cross-sections for the atoms and assumed that all heavy metals (except noble gases) and a certain fraction of C and O are in solid form. To account for different dust sizes they used self-blanketing factors.

The deviations between the \citet{Igea1999} and P{\tiny RO}D{\tiny I}M{\tiny O} cross-sections are on average  $\approx\!20\%$ with a maximum of $\approx\!30\%$. For the \citet{Bethell2011a} cross-sections we find deviations of $\approx\!10\%$ with a maximum of $\approx\!50\%$, relative to P{\tiny RO}D{\tiny I}M{\tiny O}. The agreement for the gas cross-section is remarkably good considering that we did not adapt the exact same element abundances (although in all three cases some form of depleted abundances are use).

The \citet{Bethell2011a} dust cross-sections are about a factor of $1.2-2$ higher than the P{\tiny RO}D{\tiny I}M{\tiny O} cross-sections. The differences are probably caused by the different dust properties (e.g. in the P{\tiny RO}D{\tiny I}M{\tiny O} case $a_\mathrm{max}=3000\,\mathrm{\mu m}$). However, considering the general uncertainties connected to X-ray dust opacities the agreement is quite good. 
\section{Comparison to the Monte Carlo X-ray radiative transfer code \uppercase{MOCASSIN}}
\label{sec:comperco}
\citet{Ercolano2013b} presented a comparison of the 3D Monte Carlo radiative transfer and photoionization code \texttt{MOCASSIN} \citep{Ercolano2008} to the Monte Carlo X-ray radiative transfer model of \citet{Igea1999}. For the comparison a minimum mass solar nebula disk model is used (see \citealt{Igea1999} for details). The X-ray emitting source is modelled via two rings with radius $5\,R_\sun$ located at a height of $5\,R_\sun$ above and below the star. Dust absorption and scattering is not considered in those models. \citet{Ercolano2013b} found a good agreement for the resulting X-ray ionization rates per hydrogen nucleus between the two codes.

\begin{figure}
\resizebox{\hsize}{!}{\includegraphics{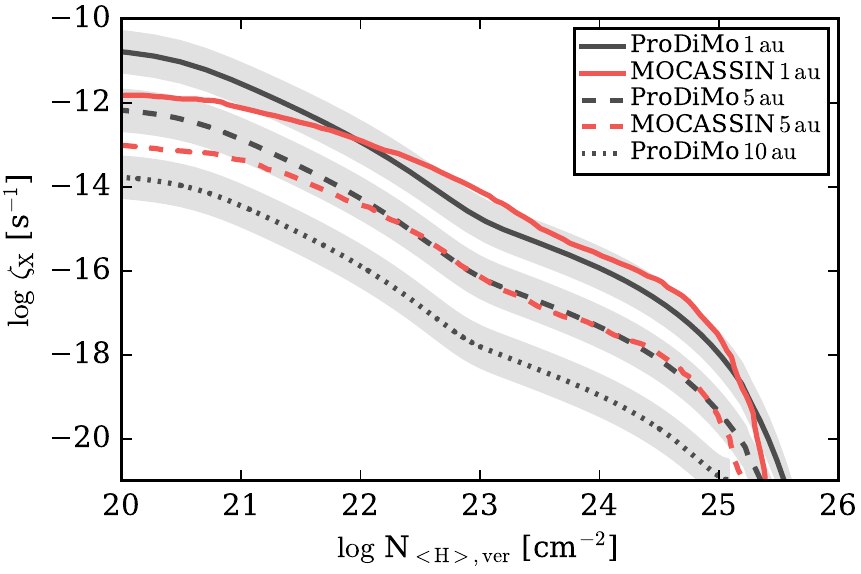}}
\caption{Comparison of the X-ray ionization rate $\zeta_\mathrm{X}$ as a function of vertical hydrogen column density $N_\mathrm{\langle H\rangle,ver}$ at radii of $1\,\mathrm{au}$ (solid lines), $5\,\mathrm{au}$ (dashed lines) and $10\,\mathrm{au}$ (dotted line). The red lines are the results from \texttt{MOCASSIN} the black lines are the results from P{\tiny RO}D{\tiny I}M{\tiny O}. The $10\,\mathrm{au}$ result from \texttt{MOCASSIN} were not available for this test case. The $10\,\mathrm{au}$ results have been scaled down by a factor of 10 for better visibility. The gray shaded area indicates a factor of three deviation relative to the P{\tiny RO}D{\tiny I}M{\tiny O} results. An isothermal X-ray spectrum with $kT_\mathrm{X}=5\,\mathrm{keV}$ and $L_\mathrm{X}=10^{29}\,\mathrm{erg\,s^{-1}}$ is used.} 
\label{fig:comperco}
\end{figure}
In Fig.~\ref{fig:comperco} we show a comparison of our X-ray radiative transfer model to the results from \texttt{MOCASSIN}. For this comparison we used the same physical disk parameters and the same initial element abundances (the ``IG99 depleted'') as given in \citet{Ercolano2013b}. The X-ray luminosity is \mbox{$L_\mathrm{X}(1-20~\mathrm{keV})=10^{29}\,\mathrm{erg\,s^{-1}}$} and the spectrum has a single temperature of $kT_\mathrm{X}=5\,\mathrm{keV}$. In contrast to the models presented in \citet{Ercolano2013b} we assume that the X-rays are emitted at the stellar surface. The X-ray ionization rate $\zeta_\mathrm{X}$ is simply calculated for the given initial abundances (i.e. all species are neutral and all hydrogen is  in molecular form) without solving for the chemistry. 

As seen from Fig.~\ref{fig:comperco}, the agreement of the two codes is quite good but there are also some significant differences. We want to emphasize that the radiative transfer methods used are very different. \texttt{MOCASSIN} uses the Monte Carlo approach with a proper treatment of anisotropic scattering. In P{\tiny RO}D{\tiny I}M{\tiny O}, we apply a ray-base method (discrete-ordinates) with a simple approximation for the anisotropic scattering. Scattering dominates $\zeta_\mathrm{X}$ for $N_\mathrm{H,ver}\gtrsim10^{23}\,\mathrm{cm^{-2}}$ (scattering shoulder \citealt{Igea1999}). In this region the results of both codes are within a factor of three, indicating that our simplified treatment of anisotropic scattering is a reasonable approximation.

The largest differences are actually at low column densities of $N_\mathrm{\langle H\rangle,ver}\lesssim10^{21}\,\mathrm{cm^{-2}}$ where $\zeta_\mathrm{X}$ is solely dominated by absorption of low energy X-ray photons (here $E_\mathrm{X}\approx 1\,\mathrm{keV}$ ). Those deviations might be caused by the exact treatment of the low energy-cut off for the X-ray spectrum. We did run a test with a cut-off at $1.1\,\mathrm{keV}$ in that case $\zeta_\mathrm{X}$ drops by a factor of a few at low column densities but is not affected at high column densities. However, we also performed a test with twice as much energy bands in our X-ray RT to more accurately treat the low energy cut-off, but we did not find significant differences. So, the origin of these deviations remains rather unclear. Although $\zeta_X$ at low columns is relevant for photo-evaporative winds \citep[e.g.][]{Ercolano2008c} it is of less relevance for our here presented models as we are mainly interested in the impact of X-ray ionization on the chemistry.

The deviations around $N_\mathrm{\langle H\rangle,ver}=10^{23}\,\mathrm{cm^{-2}}$ are likely caused by the different locations of the X-ray emitting source. Compared to a pure stellar X-ray source, the X-ray photons emitted by an elevated ring source have encountered a lower column of material when they reach a certain height of the disk \citep[see also][]{Igea1999}. At larger distances the geometrical effects (e.g. incident angle on the disk surface) are less severe. The actual location of the X-ray source in T~Tauri stars is not well known which is a general uncertainty in the models. However, as discussed by \citet{Ercolano2009b} there are observational indications for a location of the X-ray source at or very close ($\lesssim1\,R_\sun$) to the stellar surface. 
\begin{figure}
\resizebox{\hsize}{!}{\includegraphics{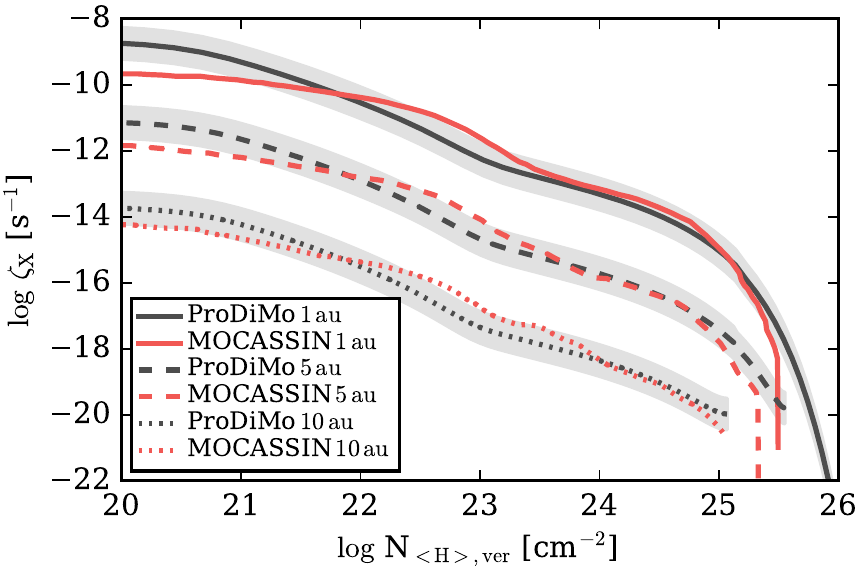}}
\caption{Same as Fig.~\ref{fig:comperco} but for a flared X-ray spectrum and low metal abundances. The 5 and $10\,\mathrm{au}$ results  have been scaled down by factor of 10 and 1000, respectively. The X-ray luminosity is $L_\mathrm{X}=2\times10^{31}\,\mathrm{erg\,s^{-1}}$ and the plasma temperature is $kT_\mathrm{X}=12\,\mathrm{keV}$ (the COUP flared spectrum in \citealt{Ercolano2013b}).} 
\label{fig:comperco2}
\end{figure}

In Fig.~\ref{fig:comperco2}, we present a similar comparison as in Fig.~\ref{fig:comperco} but for a flared X-ray spectrum with $L_\mathrm{X}(1-20~\mathrm{keV})=2\times10^{31}\,\mathrm{erg\,s^{-1}}$ and $kT_\mathrm{X}=12\,\mathrm{keV}$ and different element abundances (``ISM depleted'' from Table~1 of \citealt{Ercolano2013b}). The deviations between the models are similar to the first test case. However, now the impact of the X-ray source location is better visible. In the \texttt{MOCASSIN} models the scattering shoulder is located at slightly deeper layers than in the P{\tiny RO}D{\tiny I}M{\tiny O} models. Due to the elevated location of the X-ray source the photons experience a lower radial column of material and can penetrate to deeper layers of the disk. The differences for $N_\mathrm{\langle H\rangle,ver}>10^{25}\,\mathrm{cm^{-2}}$ are likely of technical nature. As noted by \citet{Ercolano2013b} the sudden drop in $\zeta_\mathrm{X}$ is caused by the low number of photons reaching those layers. In the P{\tiny RO}D{\tiny I}M{\tiny O} model convergence of the iterative scheme might not be reached in the high energy range where scattering is most important. This results in an overestimate of $\zeta_\mathrm{X}$ in deep and dense regions of the disk. However, in such regions, $\zeta_\mathrm{X}\lesssim10^{-19}\,\mathrm{s^{-1}}$, which is below the ionization rates expected from cosmic rays and the decay of short-lived radionuclides \citep[e.g.][]{Umebayashi2009,Cleeves2015a}.

Although there are differences concerning the X-ray ionization rate between P{\tiny RO}D{\tiny I}M{\tiny O} and \texttt{MOCASSIN}, the overall agreement is quite good. In particular, it shows that our ray-based radiative transfer method gives similar results as the Monte Carlo radiative transfer method. However, as the calculation of the \mbox{X-ray} ionization rate is a rather complex topic, a proper and more detailed benchmarking of different X-ray radiative transfer codes is desirable for a better understanding of the remaining model differences.
\section{X-ray photoelectric cross-sections for PAHs}
\label{sec:xpahcs}
\begin{figure}
\resizebox{\hsize}{!}{\includegraphics{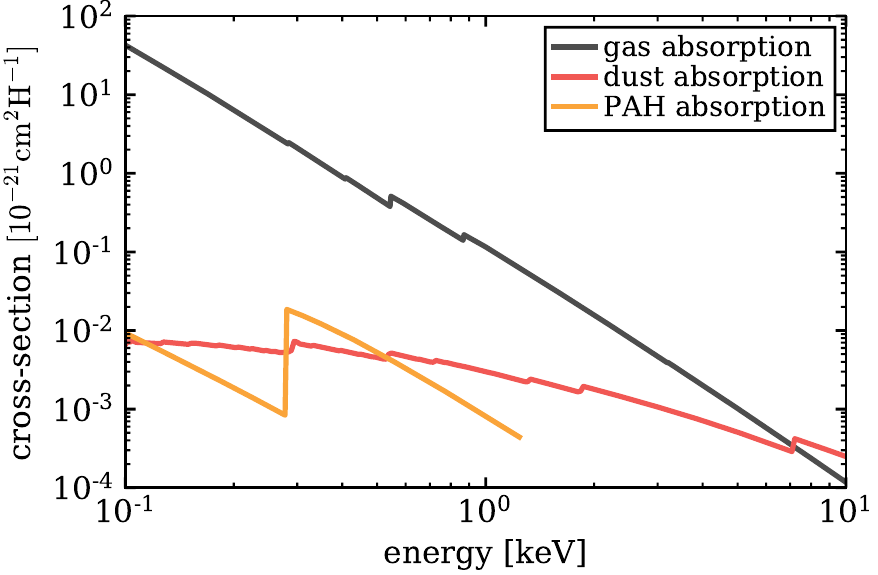}}
\caption{Comparison of PAH photoelectric (absorption) cross-sections per hydrogen nucleus in the X-ray regime to the dust and gas absorption cross-sections.}
\label{fig:xpahcs}
\end{figure}
In Fig.~\ref{fig:xpahcs} we show the photo-electric (absorption) cross-section of PAHs (Polycyclic aromatic hydrocarbons) in comparison to the X-ray gas and dust cross-sections used in our reference model (i.e. large grains dust model; see Fig.~\ref{fig:xdustopac}). The PAH absorption cross-sections are from \citet{Li2001}~\footnote[1]{\url{https://www.astro.princeton.edu/~draine/dust/dust.diel.html}}. They provide absorption cross-sections for various PAH sizes in the X-ray energy domain up to $2\,\mathrm{keV}$, assuming graphite-like optical properties. We used neutral PAH particles with 54 carbon and 18 hydrogen atoms and a radius of $\approx4.87\times10^{-4}\,\mathrm{\mu m}$ (the same PAH properties as are used for our chemical model) and considered an ISM like PAH abundance of $\approx2.8\times10^{-7}$ \citep{Tielens2008}. Further details on the treatment of PAHs in our model can be found in \citet{Woitke2016}.

It is clear from Fig.~\ref{fig:xpahcs} that, compared to the gas, PAHs play an insignificant role in the X-ray radiative transfer and for the disk X-ray radiation field. Furthermore, the PAH abundance in disks around \mbox{T Tauri} stars is likely significantly lower than in the ISM \citep[e.g.][]{Geers2006}, which further reduces the importance of PAHs as an opacity source (similar to the far-UV wavelength regime, see \citealt{Woitke2016}). For all models presented in this paper, we assumed a PAH abundance reduced by a factor of 100 compared to the ISM. Nevertheless, X-rays might play an essential role for the destruction of PAHs in disks. Due to their high energies, X-ray photons can efficiently destroy PAHs on timescales significantly shorter than typical disk lifetimes \citep{Siebenmorgen2010,Siebenmorgen2012c}. However, here we are mainly interested in the overall X-ray disk radiation field and the resulting X-ray ionization rates and therefore do not include PAHs in the X-ray radiative transfer. 
\section{HCO$^+$ formation pathway via excited molecular hydrogen}
\label{sec:HCOpHtexc}
The formation pathway for HCO$^+$ via excited molecular hydrogen $\mathrm{H_2^*}$ works in the following 
\begin{align}
\mathrm{H_2^*}+\mathrm{C^+}\rightarrow&\;\mathrm{CH^+\;\;}+\mathrm{H_2}\\ 
\mathrm{CH^+}+\mathrm{H_2}\rightarrow&\;\mathrm{CH_2^+\;\;}+\mathrm{H}\\
\mathrm{CH_2^+}+\mathrm{H_2}\rightarrow&\;\mathrm{CH_3^+\;\;} + \mathrm{H}\\
\mathrm{CH_3^+}+\mathrm{O\;}\rightarrow&\;\mathrm{HCO^+} + \mathrm{H_2}
\end{align}
Key here is the reaction $\mathrm{H_2^*+C^+}\rightarrow\mathrm{CH^+ + H_2}$, where we use the measured rate of \citet{Hierl1997} with an extrapolation to lower temperatures. However, in particular the rate at low temperatures ($T<100\,\mathrm{K}$) is uncertain, and theoretical models indicate significantly lower rates (by orders of magnitude, \citealt{Zanchet2013}). A more detailed discussion and a comparison of different rates for $\mathrm{H_2^*+C^+}$ is  presented in \citet{Kamp2017}.
\end{appendix}
\end{document}